\documentclass[pre,twocolumn]{revtex4-2}
\usepackage{amsmath}
\usepackage{amssymb}
\usepackage{graphicx}
\usepackage{braket}
\usepackage{stmaryrd}
\usepackage{hyperref}
\usepackage{color}
\usepackage{dsfont}
\usepackage{bm}
\usepackage{booktabs}

\hypersetup{
    colorlinks,
    citecolor=blue,
    linkcolor=blue,
    urlcolor=blue
}

\newcommand{\dblbrace}[1]{\llbracket #1 \rrbracket}

\begin{document}

\title{Unified speed limits in classical and quantum dynamics via temporal Fisher information}

\author{Tomohiro Nishiyama}
\email{htam0ybboh@gmail.com}
\affiliation{Independent Researcher, Tokyo 206-0003, Japan}

\author{Yoshihiko Hasegawa}
\email{hasegawa@biom.t.u-tokyo.ac.jp}
\affiliation{Department of Information and Communication Engineering, Graduate
School of Information Science and Technology, The University of Tokyo,
Tokyo 113-8656, Japan}
\date{\today}

\begin{abstract}

The importance of Fisher information is increasing in nonequilibrium thermodynamics, as it has played a fundamental role in trade-off relations such as thermodynamic uncertainty relations and speed limits. 
In this work, we investigate temporal Fisher information, which measures the temporal information content encoded in probability distributions, for both classical and quantum systems. 
We establish that temporal Fisher information is bounded from above by physical costs, such as entropy production in classical Langevin and Markov processes and the variance of interaction Hamiltonians in open quantum systems. Conversely, temporal Fisher information is bounded from below by statistical distances (e.g., the Bhattacharyya arccos distance), leading to classical and quantum speed limits that constrain the minimal time required for state transformations. 
We perform numerical simulations on two quantum dot models to validate the obtained bounds.
Our work provides a unified perspective on speed limits from the point of view of temporal Fisher information in both classical and quantum dynamics. 

\end{abstract}

\maketitle
\section{Introduction\label{sec:Introduction}}

Fisher information plays a central role in statistical inference and estimation theory. At its core, Fisher information serves as a measure of the amount of information a random variable carries about an unknown parameter of a statistical model.
It is used in many areas of statistics, such as estimation theory, hypothesis testing, and confidence interval construction. 
For instance, the inverse of the Fisher information provides a lower bound for the variance of any unbiased estimator, which is known as the Cram\'er--Rao inequality.
The importance of Fisher information is increasing in nonequilibrium thermodynamics, as it has played a fundamental role in trade-off relations such as thermodynamic uncertainty relations \cite{Barato:2015:UncRel,Gingrich:2016:TUP} and speed limits \cite{Mandelstam:1945:QSL,Margolus:1998:QSL,Deffner:2017:QSLReview,Taddei:2013:QSL,Pires:2016:GQSL}. 

\begin{figure}
    \includegraphics[width=0.75\linewidth]{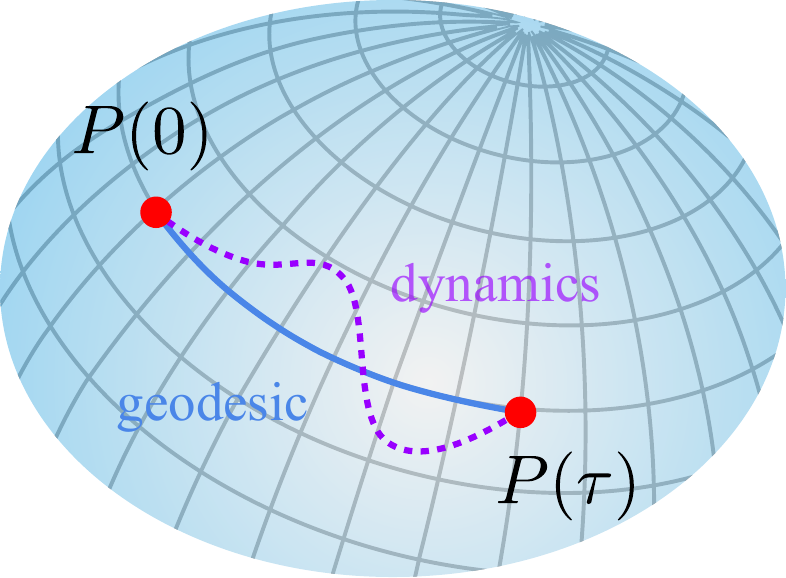}
    \caption{
Illustration of the relationship between the geodesic and the dynamics. The point $P(0)$ represents the initial position, while $P(\tau)$ represents the position after time evolution. The blue curve depicts the geodesic, the shortest path connecting $P(0)$ and $P(\tau)$. The purple dashed curve represents the trajectory of the time evolution of the system dynamics.
}
    \label{fig:geodesic}
\end{figure}

Consider the probability distribution of a stochastic process. We introduce the concept of temporal Fisher information, denoted as $\mathcal{I}_t(t)$ (cf. Eq.~\eqref{eq:def_Fisher}), which quantifies the amount of information about time contained within the probability distribution. For example, if the state described changes very little over time, it becomes difficult to determine the passage of time solely from this distribution. Therefore, temporal Fisher information measures how significantly the dynamics of the system vary with respect to time.
In a study by Wootters \cite{Wootters:1981:StatDist}, it was shown that there is a fundamental relationship between temporal Fisher information and the statistical distance (specifically, the Bhattacharyya arccos distance; see Eq.~\eqref{eq:Bhattacharyya_arccos_def}) between the initial and final states of a system. 
Specifically, the accumulated effect of temporal Fisher information over a time interval, representing the ``length'' of the trajectory traced by the system's time evolution, is always greater than or equal to the shortest possible distance between the initial and final probability distributions (Eq.~\eqref{eq:Fisher_Bhattacharyya}). This shortest distance is known as the geodesic distance in the space of probability distributions (Fig.~\ref{fig:geodesic}). In essence, the actual path taken by the system's dynamics cannot be shorter than the direct path connecting its starting and ending states.
The inequality of Eq.~\eqref{eq:Fisher_Bhattacharyya} itself represents a speed limit \cite{Ito:2018:InfoGeo,Ito:2020:TimeTURPRX}, as the distance between the initial and final states is bounded from above by an information quantity.
Indeed, temporal Fisher information is known to provide a thermodynamic length, which quantifies the distance between two equilibrium states \cite{Crooks:2007:ThermodynamicLength}. 
Moreover, temporal Fisher information provides a trade-off between time and information \cite{Nicholson:2020:TIUncRel}, which is a classical analog of the Mandelstam-Tamm speed limit \cite{Mandelstam:1945:QSL}. 
However, temporal Fisher information does not have a clear physical interpretation, preventing us from interpreting the inequality as a physical trade-off relation. 

In this manuscript, we obtain upper bounds for temporal Fisher information. 
Specifically, we obtain upper bounds for Langevin dynamics, classical Markov processes, open quantum dynamics described by joint unitary evolution on the system and environment, and non-hermitian dynamics.  
For the Langevin dynamics and Markov jump processes, we show that temporal Fisher information is bounded from above by the entropy production divided by the square of time (cf. Eq.~\eqref{eq:lambda_Langevin}). 
For open quantum dynamics, we show that temporal Fisher information is bounded from above by the variance of the interaction Hamiltonian
(cf. Eq.~\eqref{eq:general_quantum_lambda}). 
Similarly, for non-Hermitian dynamics, we show that temporal Fisher information has an upper bound comprising the variance of the dissipative components of non-Hermitian operators
(cf. Eq.~\eqref{eq:non_hermitian_lambda}). 
We demonstrate the obtained bounds through numerical simulations on two quantum dot models.
First, we consider a single quantum dot coupled to an electrode, whose dynamics in the Coulomb blockade regime is described by a two-state Markov chain. We verify the classical speed limit based on entropy production (cf. Eq.~\eqref{eq:speedlimit_Markov}) and the dynamical activity bound (cf. Eq.~\eqref{eq:CSL_dynamical_activity}).
Second, we consider a double quantum dot model, in which the left dot serves as the system and the right dot as the environment, to verify the quantum speed limit for open quantum dynamics (cf. Eq.~\eqref{eq:SL_general_open_interaction}). 
In this study, we clarify the physical meaning of temporal Fisher information and derive speed limits given by physical quantities such as entropy production.

\begin{table*}
    \centering
    \begin{tabular}{ccccc}
        \toprule
        & Langevin & Classical Markov & Open quantum & Non-Hermitian \\
        \midrule
        Upper bound & $\displaystyle  \Lambda_{\mathrm{LA}}(t):=\frac{\Sigma(t)}{2t^2}$    & $\displaystyle  \Lambda_{\mathrm{MA}}(t):=\frac{\Sigma(t)}{2t^2}$ & $\displaystyle  \Lambda_{\mathrm{OQ}}(t):=4\dblbrace{H_{SE}}(t)^2$ & $\displaystyle  \Lambda_{\mathrm{NH}}(t):=4\dblbrace{\gamma}(t)^2$\\
        Speed limit & \multicolumn{2}{c}{$\displaystyle \frac{1}{2}\int_0^\tau \sqrt{\Lambda(t)}dt\geq \mathcal{L}_P(P(0),P(\tau))$} &  \multicolumn{2}{c}{$\displaystyle \frac{1}{2}\int_0^\tau \sqrt{\Lambda(t)}dt \geq \widetilde{\mathcal{L}}_D([\rho(0)],[\rho(\tau)])$} \\
        \bottomrule
    \end{tabular}
    \caption{Summary of results. 
Upper bound $\Lambda(t)$ of temporal Fisher information and speed limits for Langevin dynamics, Markov jump processes, general open quantum dynamics, and non-Hermitian dynamics. 
$\Sigma(t)$ is the entropy production; $\dblbrace{H_{SE}}(t)$ and $\dblbrace{\gamma}(t)$ are the standard deviations of the interaction Hamiltonian and the skew-Hermitian component of the Hamiltonian, respectively. $\mathcal{L}_P$ is the Bhattacharyya arccos
distance, and $\widetilde{\mathcal{L}}_{D}$ is the unitarily residual measure of the Bures angle $\mathcal{L}_{D}$. $\rho(t)$ is a system density operator.  
}
    \label{tab:results}
\end{table*}

\section{Methods}

In this section, we present the mathematical framework that connects temporal Fisher information to speed limits in both classical and quantum dynamics. We first examine the classical case with discrete and continuous probability distributions, followed by the quantum case.

Let $P:=\{p_i\}$ and $Q:=\{q_i\}$ be discrete probability distributions. 
Consider the temporal Fisher information defined as
\begin{align}
    \mathcal{I}_t(t):=\sum_i p_i(t) (d_t \ln p_i(t))^2=-\sum_i p_i(t) d_t^2 \ln p_i(t),
    \label{eq:def_Fisher}
\end{align}
where $d_t:=d/dt$.
For continuous probability distributions $P$ and $Q$ on $\mathbb{R}^n$, we define  
$\mathcal{I}_t(t):=\int p(\mathbf{x},t) (\partial_t \ln p(\mathbf{x},t))^2 d^nx$ and $\mathcal{L}_P(P,Q):=\arccos(\int  \sqrt{p(\mathbf{x}) q(\mathbf{x})} d^nx)$.
The temporal Fisher information defined in Eq.~\eqref{eq:def_Fisher} quantifies how much information about time is encoded in the probability distribution.
Suppose that temporal Fisher information has an upper bound:
\begin{align}
    \mathcal{I}_t(t)&\le \Lambda(t),
    \label{eq:illustrative_bound}
\end{align}
where $\Lambda(t)$ is an upper bound comprising the operators determined by the dynamics (e.g., the Hamiltonian, entropy production, etc.). 
By the result of Ref.~\cite{Wootters:1981:StatDist}, the following relation holds:
\begin{align}
    \frac{1}{2}\int_0^\tau \sqrt{\mathcal{I}_t(t)}dt\geq \mathcal{L}_P(P(0),P(\tau)),
    \label{eq:Fisher_Bhattacharyya}
\end{align}
where $\mathcal{L}_P(P(0),P(\tau))$ is the Bhattacharyya arccos distance: 
\begin{align}
    \mathcal{L}_P(P,Q):=\arccos\left(\sum_i \sqrt{p_i q_i}\right).
    \label{eq:Bhattacharyya_arccos_def}
\end{align}
From Eq.~\eqref{eq:Fisher_Bhattacharyya}, we obtain the speed limit:
\begin{align}
    \frac{1}{2}\int_0^\tau \sqrt{\Lambda(t)}dt\geq \mathcal{L}_P(P(0),P(\tau)).
    \label{eq:base_SL_Bhattacharyya}
\end{align}

We next consider the quantum case. 
Although we can define classical temporal Fisher information using the eigenvalues of density operators, the Bhattacharyya arccos distance cannot be applied for the following reason. In classical dynamics, probability distributions are defined over positions $\mathbf{x}$ or discrete states $\{i\}$. However, in quantum dynamics, the correspondence between the eigenvalues of the density operators for the initial and final states cannot be determined solely from their spectral decompositions.
For example, consider the initial and final density operators with spectral decompositions
$\rho(0) = \sum_i p_i(0)\, \ket{p_i(0)}\bra{p_i(0)}$ and $\rho(\tau) = \sum_i p_i(\tau)\, \ket{p_i(\tau)}\bra{p_i(\tau)}$, respectively. 
We cannot simply compute $\sum_i \sqrt{p_i(0) p_i(\tau)}$ because there is no correspondence between $p_i(0)$ and $p_i(\tau)$ based only on the spectral decompositions. In other words, the time evolution of the eigenvalues must be known to determine their correspondence.

In Ref.~\cite{arXiv.2412.02231}, we introduced unitarily residual measures to quantify dissipation by isolating the non-unitary components of quantum dynamics.
Let $\mathfrak{M}$ be a set of density matrices and $\rho, \sigma\in\mathfrak{M}$. 
The Mandelstam-Tamm speed limit~\cite{Mandelstam:1945:QSL} is given by the Bures angle~\cite{Nielsen:2011:QuantumInfoBook}, which is a quantum generalization of the Bhattacharyya arccos distance:
\begin{align}
    \mathcal{L}_D(\rho, \sigma):= \arccos\left[\sqrt{\mathrm{Fid}(\rho, \sigma)}\right],
    \label{eq:L_D_def}
\end{align}
where $\mathrm{Fid}(\rho, \sigma)$ is the quantum fidelity: 
\begin{align}
    \mathrm{Fid}(\rho,\sigma):=\left(\mathrm{Tr}\left[\sqrt{\sqrt{\rho}\sigma\sqrt{\rho}}\right]\right)^{2}.
    \label{eq:fidelity_def}
\end{align}
To identify all quantum states that can be transitioned to via unitary transformations as a single point, we define the equivalence classes as 
\begin{align}
    [\rho]:= \{\sigma\in \mathfrak{M}:\ \sigma\sim \rho\},
    \label{eq:equivalence_class}
\end{align}
where the equivalence relation $\sim$ is defined for unitary transformations:
\begin{align}
    \rho\sim \sigma \; \text{if} \  ^{\exists} U \ \text{such that} \  U^\dagger U=\mathbb{I}, \; \sigma = U \rho U^\dagger.
    \label{eq:equivalence_relation}
\end{align}
Equation~\eqref{eq:equivalence_relation} illustrates that two states linked via a unitary transformation are regarded as equivalent.
The unitarily residual measures $\widetilde{d}$ are divergence measures between equivalence classes. Therefore, $\widetilde{d}([\rho],[\sigma])=0$ holds when $\rho$ is a unitary transformation of $\sigma$. 
The unitarily residual measures are naturally induced from quantum divergences $d(\rho,\sigma)$:
\begin{align}
    &\widetilde{d}([\rho(0)],[\rho(\tau)]):=\min_{U^{\dagger}U=V^{\dagger}V=\mathbb{I}}d(U\rho U^{\dagger},V\sigma V^{\dagger}),
    \label{eq:def_metric_quotient}
\end{align}
where the minimum is over all possible unitaries $U$ and $V$. Let $\rho=\sum_{i=1}^n p_i\ket{p_i}\bra{p_i}$ and $\sigma=\sum_{j=1}^n q_j\ket{q_j}\bra{q_j}$. Let $\mathbf{x}^\uparrow$ be a sorted vector obtained by arranging the components of $\mathbf{x}\in\mathbb{R}^n$ in non-descending order (i.e., $x_1^\uparrow\le x_2^\uparrow\le \cdots \le x_n^\uparrow$). Let $P^\uparrow$ and $Q^\uparrow$ be probability distributions whose components are $\{p^\uparrow_i\}$ and $\{q^\uparrow_i\}$, respectively. The unitarily residual measure corresponding to the Bures angle is written as the Bhattacharyya arccos distance between $P^\uparrow$ and $Q^\uparrow$:
\begin{align}
    \widetilde{\mathcal{L}}_D([\rho],[\sigma])=\mathcal{L}_P(P^\uparrow, Q^\uparrow).
    \label{eq:residual_Bures_angle}
\end{align}
Since $\sum_i a_i b_i \le \sum_i a_i^\uparrow b_i^\uparrow$ holds for the real sequences $\{a_i\}$ and $\{b_i\}$, it follows that $\mathcal{L}_P(P, Q)\geq \mathcal{L}_P(P^\uparrow, Q^\uparrow)$. 
In Eq.~\eqref{eq:def_metric_quotient}, we optimize unitary operators $U$ and $V$ to minimize the distance between states.
We note a connection to passivity in quantum thermodynamics, where passivity concerns the maximum work extractable via unitary operations (i.e., ergotropy) \cite{Allahverdyan:2004:WorkExtraction,Perarnau-Llobet:2015:ExtractableWorkCorrelations}.
Passivity is characterized by minimizing the energy expectation value of a state over all unitaries.
However, there is a fundamental difference between the unitarily residual measure and passivity.
The unitarily residual measure is introduced to compare density operators modulo unitary transformations to remove the freedom associated with unitary changes of internal dynamics.
Thus, while both concepts involve an optimization over unitary transformations, they differ in both the optimized quantity and the physical question addressed: energy minimization in the case of passivity, versus distance minimization for distinguishability in the unitarily residual measure.
Defining temporal Fisher information for eigenvalues $\{p_i(t)\}$ of density operators, we obtain the speed limits from Eq.~\eqref{eq:base_SL_Bhattacharyya}:
\begin{align}
    \frac{1}{2}\int_0^\tau \sqrt{\Lambda(t)}dt \geq \widetilde{\mathcal{L}}_D([\rho(0)],[\rho(\tau)]).
    \label{eq:base_Bhattacharyya_SL_quantum}
\end{align}
This relation is the Mandelstam-Tamm-type speed limit that focuses on the dissipative component. 
 Note that $\widetilde{\mathcal{L}}_D([\rho(0)],[\rho(\tau)])$ can be calculated from the spectral decompositions of the initial and final states. 
This contrasts with the discussion above that the Bhattacharyya arccos distance cannot be applied to the eigenvalues of density operators. 
Letting $\mathcal{P}(t):=\mathrm{Tr}[\rho(t)^2]=\sum_i p_i(t)^2$ be the purity, Eq.~\eqref{eq:base_Bhattacharyya_SL_quantum} yields the speed limit for the purity (see  Appendix~\ref{sec:purity_SL}):
\begin{align}
    2\sin\left(\frac{1}{2}\int_0^\tau \sqrt{\Lambda(t)}dt\right) \geq |\mathcal{P}(\tau)-\mathcal{P}(0)|.
    \label{eq:purity_Bhattacharyya_SL_quantum}
\end{align}

A brief remark on the expression of speed limits is in order.  
In the original formulation of the quantum speed limit \cite{Mandelstam:1945:QSL}, the bound was given as a lower bound on the time required for time evolution. 
When we define $\tau$ as the time required for time evolution, $\tau$ has the lower bound:
\begin{align}
    \tau \ge \tau_{\min},
    \label{eq:tau_taumin}
\end{align}
In Ref.~\cite{Mandelstam:1945:QSL}, $\tau$ is the time necessary for the system to transition to an orthogonal state, with the minimum time $\tau_{\min}$ defined as $\tau_{\min}=\pi/(2\dblbrace{H})$, where $\dblbrace{H}$ represents the Hamiltonian's standard deviation.
Even though Eq.~\eqref{eq:base_SL_Bhattacharyya} (and Eq.~\eqref{eq:base_Bhattacharyya_SL_quantum} as well) does not explicitly serve as a constraint on $\tau$, it can be reformulated into the structure of Eq.~\eqref{eq:tau_taumin}, a transformation frequently employed in the literature.
Specifically, Eq.~\eqref{eq:base_SL_Bhattacharyya} can be represented as
\begin{align}
    \tau\ge\tau_{\min}=\frac{2\mathcal{L}_{P}(P(0),P(\tau))}{\overline{\sqrt{\Lambda(t)}}}.
    \label{eq:Lambda_as_tau}
\end{align}
Here, $\overline{\bullet}:=\frac{1}{\tau}\int_{0}^{\tau}\bullet dt$ is the time average of the quantity over $\tau$. 
$\overline{\sqrt{\Lambda(t)}}$ can be identified as the average of ${\sqrt{\Lambda(t)}}$ over the duration $[0,\tau]$.

\section{Results}
In the previous section, we examined speed limits without detailing the underlying dynamics.
In this section, 
we show Eq.~\eqref{eq:base_SL_Bhattacharyya} for Langevin dynamics and classical Markov jump processes, and we show Eq.~\eqref{eq:base_Bhattacharyya_SL_quantum} for general open quantum dynamics and non-Hermitian dynamics in order.

\subsection{Langevin dynamics \label{subsec:Langevin}}
Consider $n$-dimensional overdamped Langevin dynamics. Let $\mathbf{x}\in\mathbb{R}^n$ be an $n$-dimensional position, and let $p(\mathbf{x},t)$ be the probability density of being in $\prod_{i=1}^n[x_i,x_i+dx_i)$ at time $t$. 
The dynamics is supposed to obey the overdamped Langevin equation:
\begin{align}
    \dot{\mathbf{x}}(t)=\mathbf{F}(\mathbf{x}(t))+\sqrt{2D}\boldsymbol{\xi}(t),
    \label{eq:Langevin_eq_def}
\end{align}
where $\mathbf{F}(\mathbf{x}(t))$ is the time-independent force, $D > 0$ is the diffusion coefficient, and $\boldsymbol{\xi}(t)$ is zero-mean Gaussian white noise with the correlation $\braket{\xi_i(t)\xi_j(t^\prime)} = \delta_{ij}\delta(t-t^\prime)$. The Boltzmann constant $k_B$ is set equal to $1$. The corresponding Fokker-Planck equation is given by
\begin{align}
    &\partial_{t}p(\mathbf{x},t)=-\nabla^\top (\boldsymbol{\nu}(\mathbf{x},t)p(\mathbf{x},t))\nonumber\\
    &=-\{\nabla^\top (\mathbf{F}(\mathbf{x})p(\mathbf{x},t))-D\Delta p(\mathbf{x},t)\},
    \label{eq:FPE_def}
\end{align}
where $\boldsymbol{\nu}(\mathbf{x},t)$ is the local mean velocity:
\begin{align}
    &\boldsymbol{\nu}(\mathbf{x},t):=\mathbf{F}(\mathbf{x})-D\nabla \ln p(\mathbf{x},t).
    \label{eq:vel_nu_def}
\end{align}
The entropy production from $t=0$ to $\tau$ is calculated as
\begin{align}
    &\Sigma(\tau)=\frac{1}{D}\int_0^\tau dt \int d^nx\boldsymbol{\nu}(\mathbf{x},t)^\top\boldsymbol{\nu}(\mathbf{x},t)  p(\mathbf{x},t).
    \label{eq:ep_Langevin}
\end{align}
We modify the force in the original system with a perturbation parameter $\theta\in\mathbb{R}$ and obtain new auxiliary dynamics: 
\begin{align}
    \mathbf{F}(\mathbf{x}, t; \theta)=\mathbf{F}(\mathbf{x}) + \theta\boldsymbol{\nu}(\mathbf{x},t).
     \label{eq:pr1}
\end{align}
We assume that the initial probability distribution  $p(\mathbf{x},0;\theta)$  is the same as in the original system. 
For infinitesimally small $\theta$, Eq.~\eqref{eq:FPE_def} is modified as 
\begin{align}
    &\partial_{t}p(\mathbf{x},t;\theta)\nonumber\\
    &=-(1+\theta)\{\nabla^\top (\mathbf{F}(\mathbf{x})p(\mathbf{x},t;\theta))-D\Delta p(\mathbf{x},t;\theta)\} +O(\theta^2).
    \label{eq:partialt_p}
\end{align}
As this equation is the time-scaled equation of Eq.~\eqref{eq:FPE_def} to first order in $\theta$ with the same initial condition, it follows that
\begin{align}
    p(\mathbf{x},t;\theta)=p(\mathbf{x},(1+\theta)t)+O(\theta^2).
    \label{eq:time_scaled_probability}
\end{align}
Let $\mathbf{\Gamma}:=\{\mathbf{x}(t)|t\in[0,\tau]\}$ be the measured trajectory and $\mathbb{P}(\mathbf{\Gamma}; \theta)$ be the path probability of $\mathbf{\Gamma}$ for the perturbation parameter $\theta$: 
\begin{align}
    \mathbb{P}(\mathbf{\Gamma}; \theta)\equiv \mathbb{P}(\mathbf{\Gamma} ;\theta\: | \mathbf{x},0)p(\mathbf{x},0)d^nx,
    \label{eq:pr2}
\end{align}
where $\mathbb{P}(\mathbf{\Gamma}; \theta \:| \mathbf{x},t)$ is the conditional probability of $\mathbf{\Gamma}$ given the position $\mathbf{x}$ at time $t$. 
The Fisher information with respect to the perturbation parameter $\theta$ for the path probability satisfies~\cite{Hasegawa:2019:CRI}
\begin{align}
    \mathfrak{I}_{\theta=0}(t):=\int \mathbb{P}(\mathbf{\Gamma}; \theta) (\partial_\theta \ln \mathbb{P}(\mathbf{\Gamma}; \theta))^2\left. \right|_{\theta=0}  \mathcal{D}\mathbf{\Gamma}=\frac{\Sigma(t)}{2},
    \label{eq:entropy_Fisher_path}
\end{align}
 where $\int \bullet \mathcal{D}\mathbf{\Gamma}$ denotes the sum over all trajectories.
The details of the derivation of this relation are shown in  
Appendix~\ref{seq:Fisher_Langevin}.
Let $\mathbf{\Gamma}(t)$ be the position of the trajectory $\mathbf{\Gamma}$ at time $t$, and let $\int_{\mathbf{\Gamma}(t)=\mathbf{x}} \bullet \mathcal{D}\mathbf{\Gamma}$ be the sum over the trajectories such that $\mathbf{\Gamma}(t)=\mathbf{x}$.  
Note that $\int_{\mathbf{\Gamma}(t)=\mathbf{x}} \mathbb{P}(\mathbf{\Gamma} ;\theta\: | \mathbf{x},t)\mathcal{D}\mathbf{\Gamma}=1$, and applying Jensen's inequality for $f(x)=x^2$ and $\mathbb{P}(\mathbf{\Gamma} ;\theta\: | \mathbf{x},t)$, we obtain
\begin{align}
     &\mathfrak{I}_{\theta}(t)=\int \mathbb{P}(\mathbf{\Gamma} ;\theta) \left(\frac{\partial_\theta  \mathbb{P}(\mathbf{\Gamma} ;\theta)}{\mathbb{P}(\mathbf{\Gamma} ;\theta)}\right)^2\mathcal{D}\mathbf{\Gamma}\nonumber\\
     &=\int d^n x\int_{\mathbf{\Gamma}(t)=\mathbf{x}}\mathcal{D}\mathbf{\Gamma}\mathbb{P}(\mathbf{\Gamma} ;\theta\: | \mathbf{x},t)p(\mathbf{x},t;\theta)\nonumber\\
     &\times \left(\frac{\partial_\theta  \left(\mathbb{P}(\mathbf{\Gamma} ;\theta\: | \mathbf{x},t)p(\mathbf{x},t;\theta)\right)}{\mathbb{P}(\mathbf{\Gamma} ;\theta\: | \mathbf{x},t)p(\mathbf{x},t;\theta)}\right)^2  \nonumber\\
     &\geq \int   p(\mathbf{x},t;\theta) \left(\frac{\partial_\theta  p(\mathbf{x},t;\theta)}{p(\mathbf{x},t;\theta)}\right)^2 d^nx.
     \label{eq:Jensen2}
\end{align} 
Equation~\eqref{eq:time_scaled_probability} and Eq.~\eqref{eq:Jensen2} yield
\begin{align}
     &\mathfrak{I}_{\theta=0}(t)\geq \int   \left.\frac{\left(\partial_\theta p(\mathbf{x},(1+\theta)t)\right)^2}{p(\mathbf{x},(1+\theta)t)}  \right|_{\theta=0}  d^nx=t^2 \mathcal{I}_t(t).
    \label{eq:path_Fisher_temporal}
\end{align}
Combining this inequality with Eq.~\eqref{eq:entropy_Fisher_path}, it follows that 
\begin{align}
    \mathcal{I}_t(t)\le \frac{\Sigma(t)}{2t^2}=:\Lambda_{\mathrm{LA}}(t),
    \label{eq:lambda_Langevin}
\end{align}
 where $\Lambda_{\mathrm{LA}}(t)$ is the upper bound (cf. Eq.~\eqref{eq:illustrative_bound}) for
the Langevin dynamics. 
From Eq.~\eqref{eq:base_SL_Bhattacharyya}, we obtain the speed limit:
\begin{align}
    \frac{1}{2\sqrt{2}}\int_0^\tau \frac{\sqrt{\Sigma(t)}}{t}dt \geq \mathcal{L}_P(P(0),P(\tau)).
    \label{eq:speedlimit_Langevin}
\end{align}
Equation~\eqref{eq:speedlimit_Langevin} is the first main result in this manuscript.

Some comments are in order regarding the derived bound. 
The second law states $\Sigma(t) \ge 0$. 
The bound of Eq.~\eqref{eq:lambda_Langevin}
(and Eq.~\eqref{eq:speedlimit_Langevin})
can be identified as a refinement of the second law, given the Bhattacharyya arccos distance between the initial and final states. 
If the Bhattacharyya arccos distance between the initial and final states is positive, then the entropy production should be positive. 
Equation~\eqref{eq:speedlimit_Langevin} is a relation in which the upper bound of the Bhattacharyya arccos distance between the initial and final states is given by the entropy production. A similar relation is known to hold for the Wasserstein distance as well \cite{Dechant:2019:Wasserstein,Ito:2024:OTReview}, which has attracted much attention in classical stochastic thermodynamics \cite{Aurell:2011:Optimal}.
It is known that the following relation holds:
\begin{align}
    \int_{0}^{\tau}\Sigma(t)dt\geq \frac{\mathcal{W}^2(P(0),P(\tau))}{D\tau},
    \label{eq:Wasserstein_CSL}
\end{align}
where $\mathcal{W}^2(P(0),P(\tau))$ is the Wasserstein distance:
\begin{align}
    \mathcal{W}^{2}\left(P_{i},P_{f}\right):=\inf_{\Pi}\int d^{n}x\int d^{n}y\|\mathbf{x}-\mathbf{y}\|^{2}\Pi(\mathbf{x},\mathbf{y}).
    \label{eq:Wasserstein_def}
\end{align}
Here, $\|\mathbf{x}-\mathbf{y}\|$ is the Euclidean distance and $\Pi(\mathbf{x},\mathbf{y})$ is the coupling function satisfying $P_{i}(\mathbf{x})=\int d^{n}y\,\Pi(\mathbf{x},\mathbf{y})$ and $P_{f}(\mathbf{y})=\int d^{n}x\,\Pi(\mathbf{x},\mathbf{y})$. 
The Wasserstein distance shown in Eq.~\eqref{eq:Wasserstein_def} is a measure that represents the distance between probability distributions and is generally known to have high computational costs. Especially in high-dimensional scenarios, the complexity can pose challenges for practical computation. Conversely, the arccos distance as described in Eq.~\eqref{eq:L_D_def} is easier to compute.
However, Eq.~\eqref{eq:speedlimit_Langevin} has limitations. For example, when considering time-dependent drift terms, it is not possible to derive speed limits from the arccos distance. In contrast, with the speed limits provided by the Wasserstein distance [Eq.~\eqref{eq:Wasserstein_CSL}], it is possible to derive speed limits even for time-dependent Langevin equations.

\subsection{Markov-jump processes \label{subsec:classical_Markov}}
Consider a continuous-time Markov jump process comprising $n$
states $\{B_1, B_2, \cdots, B_n \}$. Let $W_{ij}$ be the time-independent transition
rate from $B_j$ to $B_i$ at time $t$, and let $p_i(t)$ be the probability of being in $B_i$ at time $t$. The dynamics is supposed to obey the master equation:
\begin{align}
    \dot{p}_i(t)&=\sum_{j} W_{ij}p_j(t),
    \label{eq:master_eq_Markov}
\end{align}
where $W_{ii}=-\sum_{j(\neq i)} W_{ji}$.
Assuming the local detailed-balance condition, the entropy production is given by 
\begin{align}
    \Sigma(\tau)&=\int_0^\tau \sum_{i\neq j} W_{ij}p_j(t)\ln \frac{W_{ij}p_j(t)}{W_{ji}p_i(t)}dt .
    \label{eq:def_EP_Markov}
\end{align}
We modify the transition rate for $i\neq j$ in the original system with a perturbation parameter $\theta\in\mathbb{R}$ and obtain new auxiliary dynamics with the same initial condition as the original system: 
\begin{align}
    W_{ij}(t; \theta)=W_{ij}\left[1 + \theta \frac{W_{ij}p_j(t)-W_{ji}p_i(t)}{W_{ij}p_j(t)+W_{ji}p_i(t)}\right].
    \label{eq:def_transition_rate_theta}
\end{align}
For $i=j$, we define $W_{ii}(t; \theta)=-\sum_{j(\neq i)} W_{ji}(t; \theta)$.
For infinitesimally small $\theta$, Eq.~\eqref{eq:master_eq_Markov} is modified as
\begin{align}
    &\dot{p}_i(t; \theta)=(1+\theta)\sum_{j} W_{ij}p_j(t; \theta)+O(\theta^2).
    \label{eq:time_scaled_evolution_Markov}
\end{align}
As this equation is the time-scaled equation of Eq.~\eqref{eq:master_eq_Markov} to first order in $\theta$ with the same initial condition, it follows that $p_i(t; \theta)=p_i((1+\theta)t)+O(\theta^2)$.
We define the path probability in a similar way to the Langevin dynamics in Eq.~\eqref{eq:pr2}. The Fisher information with respect to the perturbation parameter $\theta$ for the path probability satisfies~\cite{Dechant:2019:MTUR}
\begin{align}
    \mathfrak{I}_{\theta=0}\le \frac{\Sigma(t)}{2}.
    \label{eq:entropy_Fisher_path_Markov}
\end{align}
The details of the derivation of this relation are shown in 
Appendix~\ref{seq:Fisher_Markov}.
Following a similar procedure as in Eq.~\eqref{eq:path_Fisher_temporal}, we obtain 
\begin{align}
    \mathcal{I}_t(t)\le \frac{\Sigma(t)}{2t^2}=:\Lambda_{\mathrm{MA}}(t),
    \label{eq:lambda_Markov}
\end{align}
 where $\Lambda_{\mathrm{MA}}(t)$ is the upper bound given in Eq.~\eqref{eq:illustrative_bound} for
the classical Markov jump process. 
From Eq.~\eqref{eq:base_SL_Bhattacharyya}, we obtain the speed limit:
\begin{align}
    \frac{1}{2\sqrt{2}}\int_0^\tau \frac{\sqrt{\Sigma(t)}}{t}dt \geq \mathcal{L}_P(P(0),P(\tau)).
    \label{eq:speedlimit_Markov}
\end{align}
Equations~\eqref{eq:lambda_Markov} and~\eqref{eq:speedlimit_Markov} are the same as Eqs.~\eqref{eq:lambda_Langevin} and~\eqref{eq:speedlimit_Langevin}. 
Equation~\eqref{eq:speedlimit_Markov} is the second main result in this manuscript.

The Fisher
information for the path probability also satisfies
\begin{align}
    \mathfrak{I}_{\theta=0}\le \mathcal{A}(t):=\int_0^\tau \sum_{i\neq j} W_{ij}p_j(t)\,dt.
    \label{eq:activity_Fisher_path_Markov}
\end{align}
Here, $\mathcal{A}(t)$ is the dynamical activity, which quantifies the activity of systems by the average number of jump events during $[0,\tau]$. The details of the derivation of this relation are shown in 
Appendix~\ref{seq:Fisher_Markov}.
Following a similar procedure as in Eq.~\eqref{eq:path_Fisher_temporal}, we obtain
\begin{align}
    \mathcal{I}_t(t)\le \frac{\mathcal{A}(t)}{t^2}=:\Lambda_{\mathrm{MA}}^\prime(t),
    \label{eq:LambdaMA_upperbound}
\end{align}
 where $\Lambda_{\mathrm{MA}}^\prime(t)$ is the upper bound given in Eq.~\eqref{eq:illustrative_bound} provided by the dynamical activity for the classical Markov jump process. 
From Eq.~\eqref{eq:base_SL_Bhattacharyya}, we obtain the speed limit:
\begin{align}
    \frac{1}{2}\int_0^\tau \frac{\sqrt{\mathcal{A}(t)}}{t}dt \geq \mathcal{L}_P(P(0),P(\tau)).
    \label{eq:CSL_dynamical_activity}
\end{align}
This relation was shown in Ref.~\cite{Hasegawa:2023:BulkBoundaryBoundNC}.
Dynamical activity quantifies the intensity of a system's activity. In the Langevin equation, dynamical activity diverges and therefore is not well defined.
Moreover, the bound shown in Eq.~\eqref{eq:CSL_dynamical_activity} was generalized to an open quantum scenario described by the Gorini-Kossakowski-Sudarshan-Lindblad (GKSL) equation
\cite{Gorini:1976:GKSEquation,Lindblad:1976:Generators} in Ref.~\cite{Hasegawa:2023:BulkBoundaryBoundNC}, where the dynamical activity was replaced by the quantum dynamical activity \cite{Hasegawa:2020:QTURPRL,Nishiyama:2024:ExactQDAPRE}. 
The quantum dynamical activity is known to play an important role in trade-off relations in quantum thermodynamics
\cite{Hasegawa:2020:QTURPRL,Hasegawa:2023:BulkBoundaryBoundNC,
Nakajima:2023:SLD,Nishiyama:2024:ExactQDAPRE,Nishiyama:2024:OpenQuantumRURJPA,Hasegawa:2024:ConcentrationIneqPRL}

\subsection{General open quantum dynamics \label{subseq:general_open_quantum}}
Consider a general open quantum dynamics comprising a system $S$ and an environment $E$ in an $n$-dimensional Hilbert space. 
The composite system $S+E$ evolves through a joint unitary operator $U(t)$ that acts on $\rho_{SE}(0)$. 
Then, the density operator of the composite system after the unitary evolution is
\begin{align}
    \rho_{SE}(t)= U(t)\rho_{SE}(0)U^\dagger(t).
    \label{eq:rho_SE_prime_def}
\end{align}
Let $\rho_S(t) := \mathrm{Tr}_E[\rho_{SE}(t)]$ be a system density operator, where $\mathrm{Tr}_E[\bullet]$ denotes a partial trace with respect to the 
environment. Similarly, we define $\mathrm{Tr}_S[\bullet]$ as the partial trace with respect to the system and $\mathrm{Tr}_{SE}[\bullet]:=\mathrm{Tr}_S[\mathrm{Tr}_E[\bullet]]$.
Let $H_{S}(t)$ and $H_{E}(t)$ be the Hamiltonians of $S$ and $E$. 
Let $H_{SE}(t)$ be the Hamiltonian of the system-environment interaction.
The total Hamiltonian $H(t)$ is given by
\begin{align}
    H(t):= H_{S}(t)\otimes \mathbb{I}_{E}+\mathbb{I}_{S}\otimes H_{E}(t)+H_{SE}(t),
    \label{eq:def_total_H}
\end{align}
where $\mathbb{I}_{S}$ and $\mathbb{I}_{E}$ represent the respective identity operators. 
Throughout this manuscript, we drop $\mathbb{I}_S$ and $\mathbb{I}_E$.
Taking the trace of the von Neumann equation $i\dot{\rho}_{SE}(t)=[H(t), \rho_{SE}(t)]$ with respect to the environment, the time evolution of $\rho_S(t)$ is given by 
\begin{align}
    i\dot{\rho}_S(t)=[H_S(t), \rho_S(t)] + \mathrm{Tr}_E[\,[H_{SE}(t), \rho_{SE}(t)]\,],
    \label{eq:time_evolution_rhoS}
\end{align}
where we adopt the convention of setting $\hbar =1$.
Let $\rho_S(t)=\sum_i p_i(t)\ket{p_i(t)}\bra{p_i(t)}$ and $\delta X(t):=X(t)-\mathrm{Tr}_{SE}[X(t)\rho_{SE}(t)]$ for an operator $X$.
Taking the time derivative of $p_i(t)=\braket{p_i(t)|\rho_S(t)|p_i(t)}$ and using Eq.~\eqref{eq:time_evolution_rhoS}, we obtain
\begin{align}
    &i\dot{p}_i(t)= \braket{p_i(t)|\mathrm{Tr}_E[\,[H_{SE}(t), \rho_{SE}(t)]\,]|p_i(t)}\nonumber\\
    &=\braket{p_i(t)|\mathrm{Tr}_E[\,[\delta H_{SE}(t), \rho_{SE}(t)]\,]|p_i(t)},
    \label{eq:master_eigen_general_open}
\end{align}
where we use $\braket{p_i(t)|d_t p_i(t)}+\braket{d_tp_i(t)|p_i(t)}=0$ from $\braket{p_i(t)|p_i(t)}=1$.
Let $\dblbrace{X}(t):=\sqrt{\mathrm{Tr}_{SE}[\delta X(t)^2 \rho_{SE}(t)]}$ be the standard deviation of a Hermitian 
operator $X(t)$ with respect to $\rho_{SE}(t)$.
From this equation, we obtain
\begin{align}
    \mathcal{I}_t(t)\le 4\dblbrace{H_{SE}}(t)^2=:\Lambda_{\mathrm{OQ}}(t),
    \label{eq:general_quantum_lambda}
\end{align}
 where $\Lambda_{\mathrm{OQ}}(t)$ is the upper bound given in Eq.~\eqref{eq:illustrative_bound} for the general open quantum dynamics. 
The details of the derivation of Eq.~\eqref{eq:general_quantum_lambda} are shown in 
Appendix~\ref{seq:deriv_general_open_lambda}.
From Eq.~\eqref{eq:base_Bhattacharyya_SL_quantum}, we obtain the Mandelstam-Tamm-type speed limit:
\begin{align}
    \int_0^\tau \dblbrace{H_{SE}}(t) dt\geq \widetilde{\mathcal{L}}_D([\rho_S(0)],[\rho_S(\tau)]).
    \label{eq:SL_general_open_interaction}
\end{align}
Equation~\eqref{eq:SL_general_open_interaction} is the third main result in this manuscript.
Reference~\cite{PhysRevResearch.3.023074} introduced the Mandelstam-Tamm quantum speed limit for the standard deviation of $H_S(t)+H_{SE}(t)$. Our result, as shown in Eq.~\eqref{eq:SL_general_open_interaction}, provides an upper bound that relies solely on the interaction Hamiltonian $H_{SE}(t)$.

\subsection{Non-Hermitian dynamics}
Consider the non-Hermitian dynamics governed by the non-Hermitian Hamiltonian $\mathcal{H}$.
Non-Hermitian Hamiltonians are essential in open quantum systems and nonequilibrium dynamics \cite{Ashida:2020:NonHermiteReview}. Unlike closed systems, open systems exchanging energy or particles need not preserve the norm, admitting non-Hermitian descriptions. 
A key example is the quantum trajectory approach \cite{Molmer:1993:MonteCarlo,Carmichael:2009:QuantumTrajLecture}, where a system coupled to a continuously measured environment evolves via a stochastic Schr\"odinger equation, and the no-jump dynamics can be described by non-Hermitian evolution.
Quantum speed limits for non-Hermitian dynamics have been considered in several studies 
\cite{Uzdin:2012:NonHermitianSL,Impens:2021:nonHermitianSpeedLimit,Thakuria:2024:GQSL,Nishiyama:2024:NonHermiteQSLPRA}.
Here, we obtain a speed limit from the viewpoint of temporal Fisher information. 

In general, $\mathcal{H}$ can be decomposed into
\begin{align}
    \mathcal{H}(t) = H(t) - i\gamma(t),
    \label{eq:nonHermitianHamiltonian_decompose}
\end{align}
where $H(t)$ and $\gamma(t)$ are Hermitian operators. The second term in Eq.~\eqref{eq:nonHermitianHamiltonian_decompose} is the dissipative component. Consider a density operator $\rho(t)$, whose time evolution is governed by 
\begin{align}
    i\dot{\rho}(t)=(\mathcal{H}(t)\rho(t)-\rho(t)\mathcal{H}^{\dagger}(t)).
    \label{eq:nonHermitian_Heisenberg_eq}
\end{align}
Equation~\eqref{eq:nonHermitian_Heisenberg_eq} reduces to the von Neumann equation when $\mathcal{H}(t)$ is Hermitian. 
Let $\widehat{\rho}(t)$ be a normalized density operator defined as
\begin{align}
    \widehat{\rho}(t):= \frac{\rho(t)}{\mathrm{Tr}[\rho(t)]}.
    \label{eq:normalized_density}
\end{align}
For the normalized density operator, Eq.~\eqref{eq:nonHermitian_Heisenberg_eq} is modified as 
\begin{align}
     &\dot{\widehat{\rho}}(t)=-i(\mathcal{H}(t)\widehat{\rho}(t)-\widehat{\rho}(t)\mathcal{H}^{\dagger}(t))+ 2\braket{\gamma}(t) \widehat{\rho}(t),
\label{eq:nonHermitian_Heisenberg_eq_normalized}
\end{align}
where $\braket{X}(t):=\mathrm{Tr}[X(t)\widehat{\rho}(t)]$ denotes the mean of $X(t)$.
Letting $\widehat{\rho}(t)=\sum_i p_i(t)\ket{p_i(t)}\bra{p_i(t)}$ and taking the time derivative of $p_i(t)=\braket{p_i(t)|\widehat{\rho}(t)|p_i(t)}$, we obtain
\begin{align}
    \dot{p}_i(t)= -\braket{p_i(t)|\{\delta\gamma (t), \widehat{\rho}(t)\}|p_i(t)}.
    \label{eq:master_eigen_non_hermitian}
\end{align}
Let $\dblbrace{X}(t):=\sqrt{\mathrm{Tr}[\delta X(t)^2 \widehat{\rho}(t)]}$ be the standard deviation of a Hermitian operator $X(t)$. Following a similar procedure as in Eq.~\eqref{eq:general_quantum_lambda}, we obtain 
\begin{align}
    \mathcal{I}_t(t)\le 4\dblbrace{\gamma}(t)^2=:\Lambda_{\mathrm{NH}}(t),
    \label{eq:non_hermitian_lambda}
\end{align}
 where $\Lambda_{\mathrm{NH}}(t)$ is the upper bound given in Eq.~\eqref{eq:illustrative_bound} for the non-Hermitian dynamics. 
The details of the derivation of Eq.~\eqref{eq:non_hermitian_lambda} are shown in  Appendix~\ref{seq:deriv_non_hermitian_lambda}.
From Eq.~\eqref{eq:base_Bhattacharyya_SL_quantum}, we obtain the Mandelstam-Tamm-type speed limit:
\begin{align}
    \int_0^\tau \dblbrace{\gamma}(t) dt\geq \widetilde{\mathcal{L}}_D([\widehat{\rho}(0)],[\widehat{\rho}(\tau)]).
    \label{eq:SL_non_hermitian}
\end{align}
Equations~\eqref{eq:non_hermitian_lambda} and~\eqref{eq:SL_non_hermitian} were shown in Ref.~\cite{arXiv.2412.02231}.

\section{Numerical examples}

\begin{figure}
\includegraphics[width=1\linewidth]{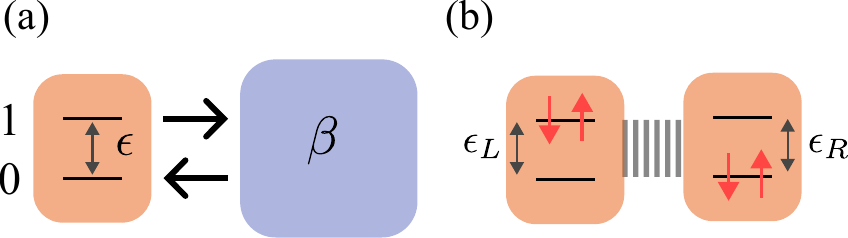}
\caption{
Illustration of quantum dot models. 
(a) 
A two-level system with energy gap $\epsilon$ is coupled to an electrode with inverse temperature $\beta$, allowing electrons to be exchanged between them.
(b)
A double quantum dot model, where the
left and right quantum dots are coupled.
The left and right systems are regarded as the system and environment, respectively. 
}
\label{fig:quantum_dot}
\end{figure}

\begin{figure}
\includegraphics[width=0.8\linewidth]{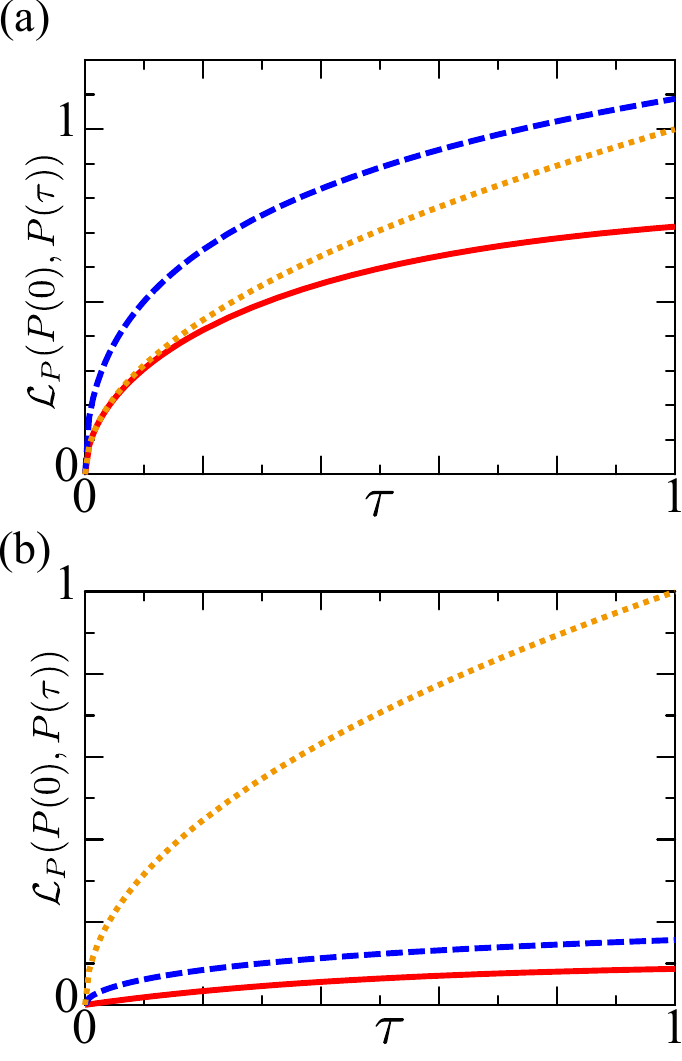}
\caption{
Numerical simulation of a
quantum dot coupled to the electrode. 
The dynamics is given by a
two-state Markov chain with transition rates $W_{10}=W_{01}=1$.
Two initial conditions are considered: (a) $P(0)=[1,0]$ and (b) $P(0)=[0.6,0.4]$.
The red solid line shows the Bhattacharyya arccos distance $\mathcal{L}_P(P(0),P(\tau))$ between the initial and final states,
the blue dotted line shows the upper bound derived from entropy production (cf. Eq.~\eqref{eq:speedlimit_Markov}),
and the orange dotted line shows the upper bound derived from dynamical activity (cf. Eq.~\eqref{eq:CSL_dynamical_activity}).
}
\label{fig:simulation_as_time}
\end{figure}

\begin{figure}
\includegraphics[width=0.8\linewidth]{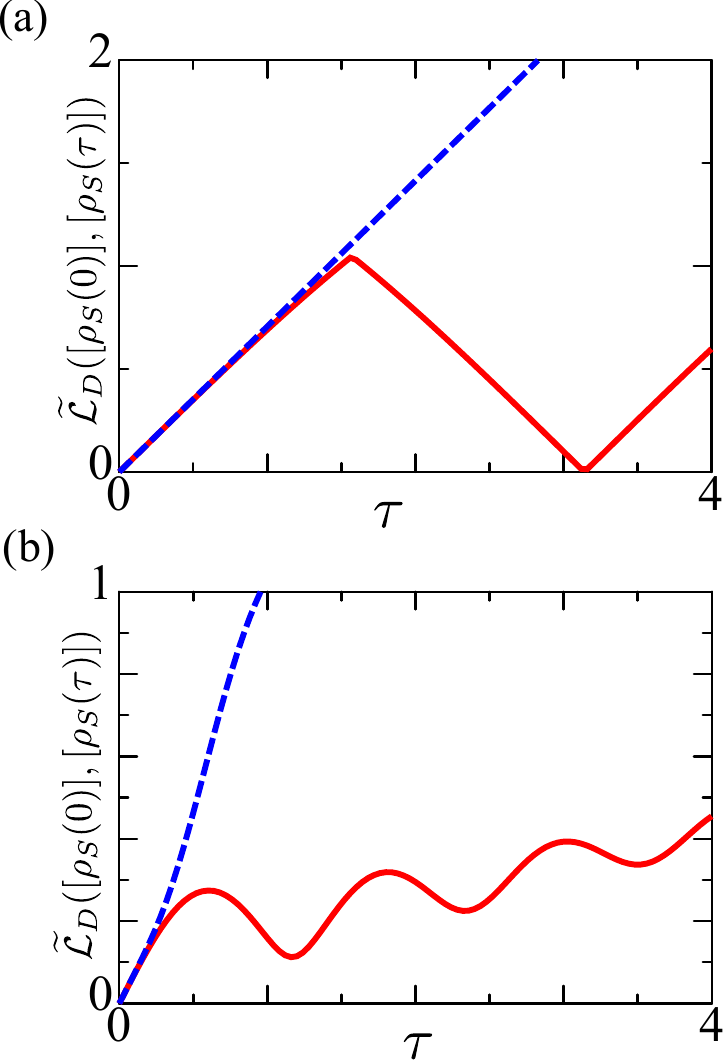}
\caption{
Numerical simulation of a double quantum dot model for two Coulomb repulsion cases: (a) the non-interacting case, $V_L=V_R=0$, and (b) the interacting case, $V_L=V_R=5$.
The red solid line shows the distance $\widetilde{\mathcal{L}}_{D}([\rho_{S}(0)],[\rho_{S}(\tau)])$ between the initial and final states, and 
the blue dotted line shows the upper bound (cf. Eq.~\eqref{eq:SL_general_open_interaction}).
The other parameters are $\epsilon_L = \epsilon_R = 1$ and $g=0.5$. 
}
\label{fig:simulation_as_time_double_qdot}
\end{figure}

We consider quantum dot models \cite{Bagrets:2003:counting,Kirsanskas:2017:QmeQ} to verify the obtained bounds, which are widely employed in stochastic and quantum thermodynamics. 

\subsection{Quantum dot coupled to an electrode}

We employ a quantum dot coupled to an electrode to verify Eq.~\eqref{eq:speedlimit_Markov}.
In this model, electrons enter and leave the quantum dot by tunneling through the tunnel barrier. A tunneling event from the electrode into the dot corresponds to a transition from the empty state to the occupied state, while the reverse process corresponds to an electron tunneling out of the dot back into the electrode.
In the regime where the Coulomb blockade is operative, the dynamics are described by the two-state Markov process:
\begin{align}
    \dot{p}_{0}(t)&=-W_{10}p_{0}(t)+W_{01}p_{1}(t),\nonumber\\\dot{p}_{1}(t)&=W_{10}p_{0}(t)-W_{01}p_{1}(t).
    \label{eq:simulation_master_equation}
\end{align}
where the transition rates are given by
\begin{align}
    W_{10}=\Gamma f(\epsilon), \quad W_{01}=\Gamma[1-f(\epsilon)].
    \label{eq:transition_rate_quantum_dots}
\end{align}
Here, $\Gamma$ denotes the tunnel coupling strength between the dot and the electrode, $\epsilon$ is the energy gap, and $f(\epsilon)$ is the Fermi-Dirac distribution.

Figure~\ref{fig:simulation_as_time} shows numerical results for this two-state Markov chain model.
The red solid line plots $\mathcal{L}_P(P(0),P(\tau))$, the distance between the initial state $P(0)$ and the final state $P(\tau)$, as a function of time $\tau$. The blue dotted line represents the entropy production upper bound from Eq.~\eqref{eq:speedlimit_Markov}, and the orange dotted line represents the dynamical activity upper bound from Eq.~\eqref{eq:CSL_dynamical_activity}. Figures~\ref{fig:simulation_as_time}(a) and (b) correspond to different initial states: $P(0)=[1,0]$ in (a) and $P(0)=[0.6, 0.4]$ in (b).
The initial state of (a) corresponds to the situation in which there is no electron in the quantum dot at $t=0$. 
The steady-state distribution is $P_\mathrm{ss}=[0.5,0.5]$, so case (b) begins closer to equilibrium.
In both cases, the upper bounds correctly constrain the distance. Moreover, comparing (a) and (b), the dynamical activity bound is tighter in (a), while the entropy production bound is tighter in (b). This demonstrates that the entropy production bound provides stronger constraints near equilibrium, whereas the dynamical activity bound becomes more effective far from equilibrium.

\subsection{Double quantum dot}

Next, we consider a double quantum dot model to verify Eq.~\eqref{eq:SL_general_open_interaction}. 
In this model, there are two quantum dots, and we identify the left dot as the system and the right dot as the environment (Fig.~\ref{fig:quantum_dot}). 
Quantum dots are often described by the Hubbard model \cite{vanderWiel:2002:doubleQuantumDots,Carrascal:2015:HubbardDimer, Wang:2024:HubbardDQDCharacterization}.
The total Hamiltonian $H$ is defined by the two-site Hubbard model:
\begin{align}
    H=H_{L}+H_{R}+H_{LR},
    \label{eq:quantum_dot_H}
\end{align}
where $H_L$ and $H_R$ are the Hamiltonians of the left and right dots, respectively:
\begin{align}
    H_{L}&=\sum_{\sigma \in \{\uparrow,\downarrow\}}\epsilon_{L}c_{L\sigma}^{\dagger}c_{L\sigma}+V_{L}c_{L\uparrow}^{\dagger}c_{L\uparrow}c_{L\downarrow}^{\dagger}c_{L\downarrow},\label{eq:HL_def}\\
    H_{R}&=\sum_{\sigma \in \{\uparrow,\downarrow\}}\epsilon_{R}c_{R\sigma}^{\dagger}c_{R\sigma}+V_{R}c_{R\uparrow}^{\dagger}c_{R\uparrow}c_{R\downarrow}^{\dagger}c_{R\downarrow}.\label{eq:HR_def}
\end{align}
Here, $c_{L\sigma}^\dagger$ is an operator that creates an electron with spin $\sigma \in \{\uparrow,\downarrow\}$ on the left dot, $\epsilon_L$ is the corresponding single-particle energy level, and $V_L$ is the on-site Coulomb repulsion energy ($c_{R\sigma}^\dagger$, $\epsilon_R$, and $V_R$ are defined analogously for the right dot).
Moreover, $H_{LR}$ in Eq.~\eqref{eq:quantum_dot_H} is the coupling Hamiltonian:
\begin{align}
    H_{LR}=-g\sum_{\sigma \in \{\uparrow,\downarrow\}}\left(c_{L\sigma}^{\dagger}c_{R\sigma}+\text{ h.c.}\right),
    \label{eq:HLR_def}
\end{align}
where $g$ is the hopping strength. 

We verify Eq.~\eqref{eq:SL_general_open_interaction} using the coupled quantum dot model.
As the initial state, we assume that both spin-$\uparrow$ and spin-$\downarrow$ electrons occupy the left quantum dot.
Figure~\ref{fig:simulation_as_time_double_qdot} shows numerical results for the double quantum dot model. 
The red solid line plots $\widetilde{\mathcal{L}}_{D}([\rho_{S}(0)],[\rho_{D}(\tau)])$, the distance between the initial system operator $\rho_S(0)$ and the final system operator $\rho_S(\tau)$, as a function of $\tau$.
The blue dashed line denotes the upper bound given in Eq.~\eqref{eq:SL_general_open_interaction}.
Figures~\ref{fig:simulation_as_time_double_qdot}(a) and (b) correspond to two different Coulomb repulsion strengths: $V_R = V_L = 0$ in (a) and $V_R = V_L = 5$ in (b), representing the non-interacting and interacting cases, respectively.
As seen from Fig.~\ref{fig:simulation_as_time_double_qdot}, 
in both cases, the upper bound correctly constrains the angle $\widetilde{\mathcal{L}}_{D}([\rho_{S}(0)],[\rho_{D}(\tau)])$. 
For the non-interacting case shown in (a), the angle exhibits a sawtooth-like behavior, and the upper bound is rather tight up to the first peak of the sawtooth. 
On the other hand, in case (b) with Coulomb repulsion, the Bures distance oscillates smoothly, and it can be seen that the upper bound is tight only in the region of small $\tau$ but quickly diverges from it.

\section{Conclusion}
In this manuscript, we have presented a link between quantum speed limits and temporal Fisher information in classical and quantum dynamics. For Langevin dynamics and classical Markov jump processes, we showed that temporal Fisher information is bounded from above by the entropy production divided by the square of time. 
For open quantum dynamics, we found that temporal Fisher information is bounded from above by the variance of the interaction Hamiltonian.
Moreover, 
through temporal Fisher information, we obtained an alternative proof of a speed limit in non-Hermitian dynamics,
which provides a unified perspective encompassing classical and open quantum dynamics. 
In addition, we derived classical and quantum speed limits from these upper bounds. 
We also performed numerical simulations on two quantum dot models to demonstrate the validity of the obtained bounds. For the single quantum dot coupled to an electrode, modeled as a two-state Markov chain, we confirmed that both the entropy production bound and the dynamical activity bound correctly constrain the Bhattacharyya arccos distance, and we observed that the entropy production bound is tighter near equilibrium, while the dynamical activity bound is more effective far from equilibrium. For the double quantum dot model, we verified the quantum speed limit for open quantum dynamics and found that the upper bound based on the variance of the interaction Hamiltonian is tight for small evolution times, though it becomes looser as the evolution time increases.
Overall, this study has contributed to a unified understanding of the quantum speed limits that govern classical and quantum dynamics.

\begin{acknowledgements}
This work was supported by the Japan Society for the Promotion of Science KAKENHI Grant Number JP23K24915.
\end{acknowledgements}

\appendix

\begin{widetext}
\section{Derivation of Eq.~\eqref{eq:purity_Bhattacharyya_SL_quantum}\label{sec:purity_SL}}
Since $|p+q-1| \le |p-1/2|+|q-1/2| \le 1$ for $p,q\in[0,1]$, it follows that  
\begin{align}
     &\sum_{i} |p^\uparrow_i - q^\uparrow_i|\geq \sum_{i} \left|(p^\uparrow_i+q^\uparrow_i-1)(p^\uparrow_i - q^\uparrow_i)\right|\geq \left| \sum_{i}(p^\uparrow_i+q^\uparrow_i-1)(p^\uparrow_i - q^\uparrow_i)\right|\geq \left|\sum_{i} p_i^2 - \sum_{i} q_i^2\right|.
     \label{eq:absolute_square}
\end{align}
Applying the Cauchy-Schwarz inequality, we obtain
\begin{align}
    &\sum_{i}|p^\uparrow_i-q^\uparrow_i|= \sum_{i}\left|\sqrt{p^\uparrow_i}-\sqrt{q^\uparrow_i}\right|\left|\sqrt{p^\uparrow_i}+\sqrt{q^\uparrow_i}\right|\le  \sqrt{\sum_{i} \left(\sqrt{p^\uparrow_i}-\sqrt{q^\uparrow_i}\right)^2\sum_{i} \left(\sqrt{p^\uparrow_i}+ \sqrt{q^\uparrow_i}\right)^2}\nonumber\\
    &= 2\sqrt{1-{\left(\sum_{i}\sqrt{p^\uparrow_iq^\uparrow_i}\right)}^2}=2\sin\left(\widetilde{\mathcal{L}}_D([\rho],[\sigma])\right),
    \label{eq:absolute_square2}
\end{align} 
where we use Eq.~\eqref{eq:residual_Bures_angle}. 
Combining this relation with Eqs.~\eqref{eq:absolute_square} and~\eqref{eq:base_Bhattacharyya_SL_quantum}, we obtain Eq.~\eqref{eq:purity_Bhattacharyya_SL_quantum}.

\section{Fisher information of path probability}
\subsection{Derivation of Eq.~\eqref{eq:entropy_Fisher_path}\label{seq:Fisher_Langevin}}
Let $p(\mathbf{x}, t+dt |\,\mathbf{y}, t;\theta)$ be the short-time transition probability density for being in position $\mathbf{x}$ 
at time $t + dt$ starting from position $\mathbf{y}$ at time $t$ with perturbation parameter $\theta$, where $dt>0$ denotes an infinitesimal time interval.
The path probability is expressed as a product of transition probabilities as 
\begin{align}
    \mathbb{P}(\mathbf{\Gamma}; \theta)=\prod_{k=1}^N \left(p(\mathbf{x}_k, t_k |\,\mathbf{x}_{k-1}, t_{k-1};\theta)d^nx_k\right) p(\mathbf{x}_0,0)d^nx_0,
    \label{eq:path_probability_product}
\end{align}
where $\tau=Ndt$ and $t_k:=kdt$.
The short-time transition probability density can be written as the Gaussian propagator~\cite{Risken:1989:FPEBook}:
\begin{align}
    p(\mathbf{x}, t+dt|\,\mathbf{y}, t;\theta)=\mathcal{N}\exp\left({-\frac{1}{4Ddt}\left(\mathbf{x}-\mathbf{y}-\mathbf{F}(\mathbf{y})dt-\theta\boldsymbol{\nu}(\mathbf{y},t)dt\right)^\top\left(\mathbf{x}-\mathbf{y}-\mathbf{F}(\mathbf{y})dt-\theta\boldsymbol{\nu}(\mathbf{y},t)dt\right) }\right),
    \label{eq:conditional_prob}
\end{align}
where $\mathcal{N}$ is a normalization factor that is independent of $\theta$ such that $\int p(\mathbf{x}, t+dt|\,\mathbf{y}, t;\theta) d^n x=1$. 
From $\mathfrak{I}_\theta(t)=\int \mathbb{P}(\mathbf{\Gamma}; \theta) (\partial_\theta \ln \mathbb{P}(\mathbf{\Gamma}; \theta))^2\mathcal{D}\mathbf{\Gamma}=-\int \mathbb{P}(\mathbf{\Gamma}; \theta) \partial_\theta^2 \ln \mathbb{P}(\mathbf{\Gamma}; \theta)\mathcal{D}\mathbf{\Gamma}$, we obtain
\begin{align}
    &\mathfrak{I}_{\theta=0}(\tau)=\sum_{l=0}^{N-1} dt \int d^nx_l \prod_{k\neq l;\ 0\le k\le N}\left(\int d^nx_{k} \right) \prod_{k\neq l+1 ;\ 1\le k\le N}p(\mathbf{x}_k, t_k |\,\mathbf{x}_{k-1}, t_{k-1}) p(\mathbf{x}_0,0)\frac{\boldsymbol{\nu}(\mathbf{x}_l,t_l)^\top \boldsymbol{\nu}(\mathbf{x}_l,t_l)}{2D} \nonumber\\
    &=\frac{1}{2D}\int_0^\tau dt \int d^nx \boldsymbol{\nu}(\mathbf{x},t)^\top \boldsymbol{\nu}(\mathbf{x},t)p(\mathbf{x}, t)=\frac{\Sigma(\tau)}{2},
    \label{eq:Fisher_information_J}
\end{align}
where $p(\mathbf{x}_k, t_k |\,\mathbf{x}_{k-1}, t_{k-1}; \theta=0)=p(\mathbf{x}_k, t_k |\,\mathbf{x}_{k-1}, t_{k-1})$.

\subsection{Derivation of Eqs.~\eqref{eq:entropy_Fisher_path_Markov} and~\eqref{eq:activity_Fisher_path_Markov}\label{seq:Fisher_Markov}}
Let $p(i, t+dt |\,j, t;\theta)$ be the short-time transition probability for being in state $B_i$ 
at time $t + dt$ starting from state $B_j$ at time $t$.
As in Eq.~\eqref{eq:path_probability_product}, we obtain
\begin{align}
    \mathbb{P}(\mathbf{\Gamma}; \theta)=\prod_{k=1}^N p(j_k, t_k |\,j_{k-1}, t_{k-1};\theta)p(j_0,0).
    \label{eq:path_parobability_Markov}
\end{align}
If a jump occurs from state $B_j$ to $B_i$ at time $t$, the transition probability is given by
$p(i, t+dt |\,j, t;\theta)=W_{ij}(t;\theta)dt$, and $p(i, t+dt |\,i, t;\theta)=1+W_{ii}(t;\theta)dt$ when no jump occurs. Combining $\mathfrak{I}_\theta(t)=-\int \mathbb{P}(\mathbf{\Gamma}; \theta) \partial_\theta^2 \ln \mathbb{P}(\mathbf{\Gamma}; \theta)\mathcal{D}\mathbf{\Gamma}$ with Eq.~\eqref{eq:def_transition_rate_theta} and $W_{ii}(t; \theta)=-\sum_{j(\neq i)} W_{ji}(t; \theta)$, we obtain
\begin{align}
    &\mathfrak{I}_{\theta=0}(\tau)=\sum_{k=1}^N \sum_{i\neq j}\int \mathcal{D}\mathbf{\Gamma} \delta_{\mathbf{\Gamma}(t_k), B_i}\delta_{\mathbf{\Gamma}(t_{k-1}), B_j} {\left(\frac{W_{ij}p_j(t_{k-1})-W_{ji}p_i(t_{k-1})}{W_{ij}p_j(t_{k-1})+W_{ji}p_i(t_{k-1})}\right)}^2 \mathbb{P}(\mathbf{\Gamma})\nonumber\\
    &=\int_0^\tau  \sum_{i> j} \frac{\left(W_{ij}p_j(t)-W_{ji}p_i(t)\right)^2}{W_{ij}p_j(t)+W_{ji}p_i(t)} dt=:\frac{1}{2}\Sigma_{\mathrm{ps}}(\tau),
    \label{eq:Fisher_pseudo_entropy}
\end{align}
where $\delta_{x,y}$ is the Kronecker delta and $\Sigma_{\mathrm{ps}}(\tau)$ is the pseudo-entropy production~\cite{Shiraishi:2016:UniversalTradeoff}.
From $2(a-b)^2/(a+b)\le (a-b)\ln(a/b)$ and $(a-b)^2/(a+b)\le a+b$ for $a,b>0$, we obtain $\Sigma(\tau)\geq \Sigma_{\mathrm{ps}}(\tau)$ and $\mathcal{A}(\tau)\geq \Sigma_{\mathrm{ps}}(\tau)/2$, respectively. Combining these relations with Eq.~\eqref{eq:Fisher_pseudo_entropy} yields Eq.~\eqref{eq:entropy_Fisher_path_Markov} and~\eqref{eq:activity_Fisher_path_Markov}.

\section{Derivation of Eq.~\eqref{eq:general_quantum_lambda}\label{seq:deriv_general_open_lambda}}
For operators $X$, $Y$, and an arbitrary real number $s$, we obtain 
\begin{align}
    &0\le \braket{p_i(t)|\mathrm{Tr}_E[(sX+iY)^\dagger (sX+iY)]|p_i(t)}\nonumber\\
    &=\braket{p_i(t)|\mathrm{Tr}_E[X^\dagger X]|p_i(t)}s^2+i\braket{p_i(t)|\mathrm{Tr}_E[(X^\dagger Y-Y^\dagger X)]|p_i(t)}s+\braket{p_i(t)|\mathrm{Tr}_E[Y^\dagger Y]|p_i(t)}.
    \label{eq:quadratic_s}
\end{align} 
Since the quadratic equation with respect to $s$ is always non-negative, it follows that
\begin{align}
    |\braket{p_i(t)|\mathrm{Tr}_E[(X^\dagger Y-Y^\dagger X)]|p_i(t)}|\le 2\sqrt{\braket{p_i(t)|\mathrm{Tr}_E[X^\dagger X]|p_i(t)}\braket{p_i(t)|\mathrm{Tr}_E[Y^\dagger Y]|p_i(t)}}.
    \label{eq:trace_cauchy_schwarz}
\end{align}
Setting $X=\sqrt{\rho_{SE}(t)}\delta H_{SE}(t)$ and $Y=\sqrt{\rho_{SE}(t)}$ in this inequality and combining it with Eq.~\eqref{eq:master_eigen_general_open}, we obtain
\begin{align}
    &\mathcal{I}_t(t)=\sum_i \frac{1}{p_i(t)}|\braket{p_i(t)|\mathrm{Tr}_E[\,[\delta H_{SE}(t),\rho_{SE}(t)]\,]|p_i(t)}|^2 \nonumber\\
    &\le 4\sum_i \braket{p_i(t)|\mathrm{Tr}_E[\delta H_{SE}(t)\rho_{SE}(t)\delta H_{SE}(t) ]|p_i(t)}=4\dblbrace{H_{SE}(t)}^2,
    \label{eq:trace_cauchy_schwarz2}
\end{align}
where we use $\braket{p_i(t)|\mathrm{Tr}_E[\rho_{SE}(t)]|p_i(t)}=p_i$. 

\section{Derivation of Eq.~\eqref{eq:non_hermitian_lambda}\label{seq:deriv_non_hermitian_lambda}}

Following a procedure similar to that of Eq.~\eqref{eq:quadratic_s} for $sX+Y$, we obtain 
\begin{align}
    |\braket{p_i(t)|(X^\dagger Y+Y^\dagger X)|p_i(t)}|\le 2\sqrt{\braket{p_i(t)|X^\dagger X|p_i(t)}\braket{p_i(t)|Y^\dagger Y|p_i(t)}}.
    \label{eq:trace_cauchy_schwarz_anti}
\end{align}
Setting $X=\sqrt{\widehat{\rho}(t)}\delta\gamma (t)$ and $Y=\sqrt{\widehat{\rho}(t)}$ in this inequality and combining it with Eq.~\eqref{eq:master_eigen_non_hermitian}, we obtain
\begin{align}
    \mathcal{I}_t(t)=\sum_i \frac{1}{p_i(t)} |\braket{p_i(t)|\{\delta\gamma (t), \widehat{\rho}(t)\}|p_i(t)}|^2\le 4\sum_i \braket{p_i(t)|\delta\gamma (t)\widehat{\rho}(t)\delta\gamma (t)|p_i(t)}=4\dblbrace{\gamma}(t)^2,
    \label{eq:Fisher_information_It}
\end{align}
where we use $\braket{p_i(t)|\widehat{\rho}(t)|p_i(t)}=p_i(t)$.
\end{widetext}


\begin{thebibliography}{44}%
\makeatletter
\providecommand \@ifxundefined [1]{%
 \@ifx{#1\undefined}
}%
\providecommand \@ifnum [1]{%
 \ifnum #1\expandafter \@firstoftwo
 \else \expandafter \@secondoftwo
 \fi
}%
\providecommand \@ifx [1]{%
 \ifx #1\expandafter \@firstoftwo
 \else \expandafter \@secondoftwo
 \fi
}%
\providecommand \natexlab [1]{#1}%
\providecommand \enquote  [1]{``#1''}%
\providecommand \bibnamefont  [1]{#1}%
\providecommand \bibfnamefont [1]{#1}%
\providecommand \citenamefont [1]{#1}%
\providecommand \href@noop [0]{\@secondoftwo}%
\providecommand \href [0]{\begingroup \@sanitize@url \@href}%
\providecommand \@href[1]{\@@startlink{#1}\@@href}%
\providecommand \@@href[1]{\endgroup#1\@@endlink}%
\providecommand \@sanitize@url [0]{\catcode `\\12\catcode `\$12\catcode `\&12\catcode `\#12\catcode `\^12\catcode `\_12\catcode `\%12\relax}%
\providecommand \@@startlink[1]{}%
\providecommand \@@endlink[0]{}%
\providecommand \url  [0]{\begingroup\@sanitize@url \@url }%
\providecommand \@url [1]{\endgroup\@href {#1}{\urlprefix }}%
\providecommand \urlprefix  [0]{URL }%
\providecommand \Eprint [0]{\href }%
\providecommand \doibase [0]{https://doi.org/}%
\providecommand \selectlanguage [0]{\@gobble}%
\providecommand \bibinfo  [0]{\@secondoftwo}%
\providecommand \bibfield  [0]{\@secondoftwo}%
\providecommand \translation [1]{[#1]}%
\providecommand \BibitemOpen [0]{}%
\providecommand \bibitemStop [0]{}%
\providecommand \bibitemNoStop [0]{.\EOS\space}%
\providecommand \EOS [0]{\spacefactor3000\relax}%
\providecommand \BibitemShut  [1]{\csname bibitem#1\endcsname}%
\let\auto@bib@innerbib\@empty
\bibitem [{\citenamefont {Barato}\ and\ \citenamefont {Seifert}(2015)}]{Barato:2015:UncRel}%
  \BibitemOpen
  \bibfield  {author} {\bibinfo {author} {\bibfnamefont {A.~C.}\ \bibnamefont {Barato}}\ and\ \bibinfo {author} {\bibfnamefont {U.}~\bibnamefont {Seifert}},\ }\bibfield  {title} {\bibinfo {title} {Thermodynamic uncertainty relation for biomolecular processes},\ }\href {https://doi.org/10.1103/PhysRevLett.114.158101} {\bibfield  {journal} {\bibinfo  {journal} {Phys. Rev. Lett.}\ }\textbf {\bibinfo {volume} {114}},\ \bibinfo {pages} {158101} (\bibinfo {year} {2015})}\BibitemShut {NoStop}%
\bibitem [{\citenamefont {Gingrich}\ \emph {et~al.}(2016)\citenamefont {Gingrich}, \citenamefont {Horowitz}, \citenamefont {Perunov},\ and\ \citenamefont {England}}]{Gingrich:2016:TUP}%
  \BibitemOpen
  \bibfield  {author} {\bibinfo {author} {\bibfnamefont {T.~R.}\ \bibnamefont {Gingrich}}, \bibinfo {author} {\bibfnamefont {J.~M.}\ \bibnamefont {Horowitz}}, \bibinfo {author} {\bibfnamefont {N.}~\bibnamefont {Perunov}},\ and\ \bibinfo {author} {\bibfnamefont {J.~L.}\ \bibnamefont {England}},\ }\bibfield  {title} {\bibinfo {title} {Dissipation bounds all steady-state current fluctuations},\ }\href {https://doi.org/10.1103/PhysRevLett.116.120601} {\bibfield  {journal} {\bibinfo  {journal} {Phys. Rev. Lett.}\ }\textbf {\bibinfo {volume} {116}},\ \bibinfo {pages} {120601} (\bibinfo {year} {2016})}\BibitemShut {NoStop}%
\bibitem [{\citenamefont {Mandelstam}\ and\ \citenamefont {Tamm}(1945)}]{Mandelstam:1945:QSL}%
  \BibitemOpen
  \bibfield  {author} {\bibinfo {author} {\bibfnamefont {L.}~\bibnamefont {Mandelstam}}\ and\ \bibinfo {author} {\bibfnamefont {I.}~\bibnamefont {Tamm}},\ }\bibfield  {title} {\bibinfo {title} {The uncertainty relation between energy and time in non-relativistic quantum mechanics},\ }\href {https://doi.org/10.1007/978-3-642-74626-0_8} {\bibfield  {journal} {\bibinfo  {journal} {J. Phys. USSR}\ }\textbf {\bibinfo {volume} {9}},\ \bibinfo {pages} {249} (\bibinfo {year} {1945})}\BibitemShut {NoStop}%
\bibitem [{\citenamefont {Margolus}\ and\ \citenamefont {Levitin}(1998)}]{Margolus:1998:QSL}%
  \BibitemOpen
  \bibfield  {author} {\bibinfo {author} {\bibfnamefont {N.}~\bibnamefont {Margolus}}\ and\ \bibinfo {author} {\bibfnamefont {L.~B.}\ \bibnamefont {Levitin}},\ }\bibfield  {title} {\bibinfo {title} {The maximum speed of dynamical evolution},\ }\href {https://doi.org/10.1016/S0167-2789(98)00054-2} {\bibfield  {journal} {\bibinfo  {journal} {Physica D: Nonlinear Phenomena}\ }\textbf {\bibinfo {volume} {120}},\ \bibinfo {pages} {188 } (\bibinfo {year} {1998})}\BibitemShut {NoStop}%
\bibitem [{\citenamefont {Deffner}\ and\ \citenamefont {Campbell}(2017)}]{Deffner:2017:QSLReview}%
  \BibitemOpen
  \bibfield  {author} {\bibinfo {author} {\bibfnamefont {S.}~\bibnamefont {Deffner}}\ and\ \bibinfo {author} {\bibfnamefont {S.}~\bibnamefont {Campbell}},\ }\bibfield  {title} {\bibinfo {title} {Quantum speed limits: from {Heisenberg}'s uncertainty principle to optimal quantum control},\ }\href {https://doi.org/10.1088/1751-8121/aa86c6} {\bibfield  {journal} {\bibinfo  {journal} {J. Phys. A: Math. Theor.}\ }\textbf {\bibinfo {volume} {50}},\ \bibinfo {pages} {453001} (\bibinfo {year} {2017})}\BibitemShut {NoStop}%
\bibitem [{\citenamefont {Taddei}\ \emph {et~al.}(2013)\citenamefont {Taddei}, \citenamefont {Escher}, \citenamefont {Davidovich},\ and\ \citenamefont {de~Matos~Filho}}]{Taddei:2013:QSL}%
  \BibitemOpen
  \bibfield  {author} {\bibinfo {author} {\bibfnamefont {M.~M.}\ \bibnamefont {Taddei}}, \bibinfo {author} {\bibfnamefont {B.~M.}\ \bibnamefont {Escher}}, \bibinfo {author} {\bibfnamefont {L.}~\bibnamefont {Davidovich}},\ and\ \bibinfo {author} {\bibfnamefont {R.~L.}\ \bibnamefont {de~Matos~Filho}},\ }\bibfield  {title} {\bibinfo {title} {Quantum speed limit for physical processes},\ }\href {https://link.aps.org/doi/10.1103/PhysRevLett.110.050402} {\bibfield  {journal} {\bibinfo  {journal} {Phys. Rev. Lett.}\ }\textbf {\bibinfo {volume} {110}},\ \bibinfo {pages} {050402} (\bibinfo {year} {2013})}\BibitemShut {NoStop}%
\bibitem [{\citenamefont {Pires}\ \emph {et~al.}(2016)\citenamefont {Pires}, \citenamefont {Cianciaruso}, \citenamefont {C\'eleri}, \citenamefont {Adesso},\ and\ \citenamefont {Soares-Pinto}}]{Pires:2016:GQSL}%
  \BibitemOpen
  \bibfield  {author} {\bibinfo {author} {\bibfnamefont {D.~P.}\ \bibnamefont {Pires}}, \bibinfo {author} {\bibfnamefont {M.}~\bibnamefont {Cianciaruso}}, \bibinfo {author} {\bibfnamefont {L.~C.}\ \bibnamefont {C\'eleri}}, \bibinfo {author} {\bibfnamefont {G.}~\bibnamefont {Adesso}},\ and\ \bibinfo {author} {\bibfnamefont {D.~O.}\ \bibnamefont {Soares-Pinto}},\ }\bibfield  {title} {\bibinfo {title} {Generalized geometric quantum speed limits},\ }\href {https://doi.org/10.1103/PhysRevX.6.021031} {\bibfield  {journal} {\bibinfo  {journal} {Phys. Rev. X}\ }\textbf {\bibinfo {volume} {6}},\ \bibinfo {pages} {021031} (\bibinfo {year} {2016})}\BibitemShut {NoStop}%
\bibitem [{\citenamefont {Wootters}(1981)}]{Wootters:1981:StatDist}%
  \BibitemOpen
  \bibfield  {author} {\bibinfo {author} {\bibfnamefont {W.~K.}\ \bibnamefont {Wootters}},\ }\bibfield  {title} {\bibinfo {title} {Statistical distance and {Hilbert} space},\ }\href {https://doi.org/10.1103/PhysRevD.23.357} {\bibfield  {journal} {\bibinfo  {journal} {Phys. Rev. D}\ }\textbf {\bibinfo {volume} {23}},\ \bibinfo {pages} {357} (\bibinfo {year} {1981})}\BibitemShut {NoStop}%
\bibitem [{\citenamefont {Ito}(2018)}]{Ito:2018:InfoGeo}%
  \BibitemOpen
  \bibfield  {author} {\bibinfo {author} {\bibfnamefont {S.}~\bibnamefont {Ito}},\ }\bibfield  {title} {\bibinfo {title} {Stochastic thermodynamic interpretation of information geometry},\ }\href {https://doi.org/10.1103/PhysRevLett.121.030605} {\bibfield  {journal} {\bibinfo  {journal} {Phys. Rev. Lett.}\ }\textbf {\bibinfo {volume} {121}},\ \bibinfo {pages} {030605} (\bibinfo {year} {2018})}\BibitemShut {NoStop}%
\bibitem [{\citenamefont {Ito}\ and\ \citenamefont {Dechant}(2020)}]{Ito:2020:TimeTURPRX}%
  \BibitemOpen
  \bibfield  {author} {\bibinfo {author} {\bibfnamefont {S.}~\bibnamefont {Ito}}\ and\ \bibinfo {author} {\bibfnamefont {A.}~\bibnamefont {Dechant}},\ }\bibfield  {title} {\bibinfo {title} {Stochastic time evolution, information geometry, and the {Cram\'er}-{Rao} bound},\ }\href {https://doi.org/10.1103/PhysRevX.10.021056} {\bibfield  {journal} {\bibinfo  {journal} {Phys. Rev. X}\ }\textbf {\bibinfo {volume} {10}},\ \bibinfo {pages} {021056} (\bibinfo {year} {2020})}\BibitemShut {NoStop}%
\bibitem [{\citenamefont {Crooks}(2007)}]{Crooks:2007:ThermodynamicLength}%
  \BibitemOpen
  \bibfield  {author} {\bibinfo {author} {\bibfnamefont {G.~E.}\ \bibnamefont {Crooks}},\ }\bibfield  {title} {\bibinfo {title} {Measuring thermodynamic length},\ }\href {https://doi.org/10.1103/PhysRevLett.99.100602} {\bibfield  {journal} {\bibinfo  {journal} {Phys. Rev. Lett.}\ }\textbf {\bibinfo {volume} {99}},\ \bibinfo {pages} {100602} (\bibinfo {year} {2007})}\BibitemShut {NoStop}%
\bibitem [{\citenamefont {Nicholson}\ \emph {et~al.}(2020)\citenamefont {Nicholson}, \citenamefont {Garcia-Pintos}, \citenamefont {del Campo},\ and\ \citenamefont {Green}}]{Nicholson:2020:TIUncRel}%
  \BibitemOpen
  \bibfield  {author} {\bibinfo {author} {\bibfnamefont {S.~B.}\ \bibnamefont {Nicholson}}, \bibinfo {author} {\bibfnamefont {L.~P.}\ \bibnamefont {Garcia-Pintos}}, \bibinfo {author} {\bibfnamefont {A.}~\bibnamefont {del Campo}},\ and\ \bibinfo {author} {\bibfnamefont {J.~R.}\ \bibnamefont {Green}},\ }\bibfield  {title} {\bibinfo {title} {Time-information uncertainty relations in thermodynamics},\ }\href {https://doi.org/10.1038/s41567-020-0981-y} {\bibfield  {journal} {\bibinfo  {journal} {Nat. Phys.}\ }\textbf {\bibinfo {volume} {16}},\ \bibinfo {pages} {1211} (\bibinfo {year} {2020})}\BibitemShut {NoStop}%
\bibitem [{\citenamefont {Nishiyama}\ and\ \citenamefont {Hasegawa}(2024{\natexlab{a}})}]{arXiv.2412.02231}%
  \BibitemOpen
  \bibfield  {author} {\bibinfo {author} {\bibfnamefont {T.}~\bibnamefont {Nishiyama}}\ and\ \bibinfo {author} {\bibfnamefont {Y.}~\bibnamefont {Hasegawa}},\ }\bibfield  {title} {\bibinfo {title} {A unified framework of unitarily residual measures for quantifying dissipation},\ }\href {https://doi.org/10.48550/arXiv.2412.02231} {\bibfield  {journal} {\bibinfo  {journal} {arXiv:2412.02231}\ } (\bibinfo {year} {2024}{\natexlab{a}})}\BibitemShut {NoStop}%
\bibitem [{\citenamefont {Nielsen}\ and\ \citenamefont {Chuang}(2011)}]{Nielsen:2011:QuantumInfoBook}%
  \BibitemOpen
  \bibfield  {author} {\bibinfo {author} {\bibfnamefont {M.~A.}\ \bibnamefont {Nielsen}}\ and\ \bibinfo {author} {\bibfnamefont {I.~L.}\ \bibnamefont {Chuang}},\ }\href {https://doi.org/10.1017/CBO9780511976667} {\emph {\bibinfo {title} {Quantum Computation and Quantum Information}}}\ (\bibinfo  {publisher} {Cambridge University Press},\ \bibinfo {address} {New York, NY, USA},\ \bibinfo {year} {2011})\BibitemShut {NoStop}%
\bibitem [{\citenamefont {Allahverdyan}\ \emph {et~al.}(2004)\citenamefont {Allahverdyan}, \citenamefont {Balian},\ and\ \citenamefont {Nieuwenhuizen}}]{Allahverdyan:2004:WorkExtraction}%
  \BibitemOpen
  \bibfield  {author} {\bibinfo {author} {\bibfnamefont {A.~E.}\ \bibnamefont {Allahverdyan}}, \bibinfo {author} {\bibfnamefont {R.}~\bibnamefont {Balian}},\ and\ \bibinfo {author} {\bibfnamefont {T.~M.}\ \bibnamefont {Nieuwenhuizen}},\ }\bibfield  {title} {\bibinfo {title} {Maximal work extraction from finite quantum systems},\ }\href {https://doi.org/10.1209/epl/i2004-10101-2} {\bibfield  {journal} {\bibinfo  {journal} {EPL}\ }\textbf {\bibinfo {volume} {67}},\ \bibinfo {pages} {565} (\bibinfo {year} {2004})}\BibitemShut {NoStop}%
\bibitem [{\citenamefont {Perarnau-Llobet}\ \emph {et~al.}(2015)\citenamefont {Perarnau-Llobet}, \citenamefont {Hovhannisyan}, \citenamefont {Huber}, \citenamefont {Skrzypczyk}, \citenamefont {Brunner},\ and\ \citenamefont {Ac{\'{\i}}n}}]{Perarnau-Llobet:2015:ExtractableWorkCorrelations}%
  \BibitemOpen
  \bibfield  {author} {\bibinfo {author} {\bibfnamefont {M.}~\bibnamefont {Perarnau-Llobet}}, \bibinfo {author} {\bibfnamefont {K.~V.}\ \bibnamefont {Hovhannisyan}}, \bibinfo {author} {\bibfnamefont {M.}~\bibnamefont {Huber}}, \bibinfo {author} {\bibfnamefont {P.}~\bibnamefont {Skrzypczyk}}, \bibinfo {author} {\bibfnamefont {N.}~\bibnamefont {Brunner}},\ and\ \bibinfo {author} {\bibfnamefont {A.}~\bibnamefont {Ac{\'{\i}}n}},\ }\bibfield  {title} {\bibinfo {title} {Extractable work from correlations},\ }\href {https://doi.org/10.1103/PhysRevX.5.041011} {\bibfield  {journal} {\bibinfo  {journal} {Phys. Rev. X}\ }\textbf {\bibinfo {volume} {5}},\ \bibinfo {pages} {041011} (\bibinfo {year} {2015})}\BibitemShut {NoStop}%
\bibitem [{\citenamefont {Hasegawa}\ and\ \citenamefont {Van~Vu}(2019)}]{Hasegawa:2019:CRI}%
  \BibitemOpen
  \bibfield  {author} {\bibinfo {author} {\bibfnamefont {Y.}~\bibnamefont {Hasegawa}}\ and\ \bibinfo {author} {\bibfnamefont {T.}~\bibnamefont {Van~Vu}},\ }\bibfield  {title} {\bibinfo {title} {Uncertainty relations in stochastic processes: An information inequality approach},\ }\href {https://doi.org/10.1103/PhysRevE.99.062126} {\bibfield  {journal} {\bibinfo  {journal} {Phys. Rev. E}\ }\textbf {\bibinfo {volume} {99}},\ \bibinfo {pages} {062126} (\bibinfo {year} {2019})}\BibitemShut {NoStop}%
\bibitem [{\citenamefont {Dechant}\ and\ \citenamefont {Sakurai}(2019)}]{Dechant:2019:Wasserstein}%
  \BibitemOpen
  \bibfield  {author} {\bibinfo {author} {\bibfnamefont {A.}~\bibnamefont {Dechant}}\ and\ \bibinfo {author} {\bibfnamefont {Y.}~\bibnamefont {Sakurai}},\ }\bibfield  {title} {\bibinfo {title} {Thermodynamic interpretation of {Wasserstein} distance},\ }\href {https://arxiv.org/abs/1912.08405} {\bibfield  {journal} {\bibinfo  {journal} {arXiv:1912.08405}\ } (\bibinfo {year} {2019})}\BibitemShut {NoStop}%
\bibitem [{\citenamefont {Ito}(2024)}]{Ito:2024:OTReview}%
  \BibitemOpen
  \bibfield  {author} {\bibinfo {author} {\bibfnamefont {S.}~\bibnamefont {Ito}},\ }\bibfield  {title} {\bibinfo {title} {Geometric thermodynamics for the fokker--planck equation: stochastic thermodynamic links between information geometry and optimal transport},\ }\href {https://doi.org/10.1007/s41884-023-00102-3} {\bibfield  {journal} {\bibinfo  {journal} {Inf. Geom.}\ }\textbf {\bibinfo {volume} {7}},\ \bibinfo {pages} {441} (\bibinfo {year} {2024})}\BibitemShut {NoStop}%
\bibitem [{\citenamefont {Aurell}\ \emph {et~al.}(2011)\citenamefont {Aurell}, \citenamefont {Mej{\'{\i}}a-Monasterio},\ and\ \citenamefont {Muratore-Ginanneschi}}]{Aurell:2011:Optimal}%
  \BibitemOpen
  \bibfield  {author} {\bibinfo {author} {\bibfnamefont {E.}~\bibnamefont {Aurell}}, \bibinfo {author} {\bibfnamefont {C.}~\bibnamefont {Mej{\'{\i}}a-Monasterio}},\ and\ \bibinfo {author} {\bibfnamefont {P.}~\bibnamefont {Muratore-Ginanneschi}},\ }\bibfield  {title} {\bibinfo {title} {Optimal protocols and optimal transport in stochastic thermodynamics},\ }\href {https://doi.org/10.1103/PhysRevLett.106.250601} {\bibfield  {journal} {\bibinfo  {journal} {Phys. Rev. Lett.}\ }\textbf {\bibinfo {volume} {106}},\ \bibinfo {pages} {250601} (\bibinfo {year} {2011})}\BibitemShut {NoStop}%
\bibitem [{\citenamefont {Dechant}(2019)}]{Dechant:2019:MTUR}%
  \BibitemOpen
  \bibfield  {author} {\bibinfo {author} {\bibfnamefont {A.}~\bibnamefont {Dechant}},\ }\bibfield  {title} {\bibinfo {title} {Multidimensional thermodynamic uncertainty relations},\ }\href {https://doi.org/10.1088%2F1751-8121%2Faaf3ff} {\bibfield  {journal} {\bibinfo  {journal} {J. Phys. A: Math. Theor.}\ }\textbf {\bibinfo {volume} {52}},\ \bibinfo {pages} {035001} (\bibinfo {year} {2019})}\BibitemShut {NoStop}%
\bibitem [{\citenamefont {Hasegawa}(2023)}]{Hasegawa:2023:BulkBoundaryBoundNC}%
  \BibitemOpen
  \bibfield  {author} {\bibinfo {author} {\bibfnamefont {Y.}~\bibnamefont {Hasegawa}},\ }\bibfield  {title} {\bibinfo {title} {Unifying speed limit, thermodynamic uncertainty relation and {Heisenberg} principle via bulk-boundary correspondence},\ }\href {https://doi.org/10.1038/s41467-023-38074-8} {\bibfield  {journal} {\bibinfo  {journal} {Nat. Commun.}\ }\textbf {\bibinfo {volume} {14}},\ \bibinfo {pages} {2828} (\bibinfo {year} {2023})}\BibitemShut {NoStop}%
\bibitem [{\citenamefont {Gorini}\ \emph {et~al.}(1976)\citenamefont {Gorini}, \citenamefont {Kossakowski},\ and\ \citenamefont {Sudarshan}}]{Gorini:1976:GKSEquation}%
  \BibitemOpen
  \bibfield  {author} {\bibinfo {author} {\bibfnamefont {V.}~\bibnamefont {Gorini}}, \bibinfo {author} {\bibfnamefont {A.}~\bibnamefont {Kossakowski}},\ and\ \bibinfo {author} {\bibfnamefont {E.~C.~G.}\ \bibnamefont {Sudarshan}},\ }\bibfield  {title} {\bibinfo {title} {{Completely positive dynamical semigroups of {N}‐level systems}},\ }\href {https://doi.org/10.1063/1.522979} {\bibfield  {journal} {\bibinfo  {journal} {J. Math. Phys.}\ }\textbf {\bibinfo {volume} {17}},\ \bibinfo {pages} {821} (\bibinfo {year} {1976})}\BibitemShut {NoStop}%
\bibitem [{\citenamefont {Lindblad}(1976)}]{Lindblad:1976:Generators}%
  \BibitemOpen
  \bibfield  {author} {\bibinfo {author} {\bibfnamefont {G.}~\bibnamefont {Lindblad}},\ }\bibfield  {title} {\bibinfo {title} {On the generators of quantum dynamical semigroups},\ }\href {https://doi.org/10.1007/BF01608499} {\bibfield  {journal} {\bibinfo  {journal} {Commun. Math. Phys.}\ }\textbf {\bibinfo {volume} {48}},\ \bibinfo {pages} {119} (\bibinfo {year} {1976})}\BibitemShut {NoStop}%
\bibitem [{\citenamefont {Hasegawa}(2020)}]{Hasegawa:2020:QTURPRL}%
  \BibitemOpen
  \bibfield  {author} {\bibinfo {author} {\bibfnamefont {Y.}~\bibnamefont {Hasegawa}},\ }\bibfield  {title} {\bibinfo {title} {Quantum thermodynamic uncertainty relation for continuous measurement},\ }\href {https://doi.org/10.1103/PhysRevLett.125.050601} {\bibfield  {journal} {\bibinfo  {journal} {Phys. Rev. Lett.}\ }\textbf {\bibinfo {volume} {125}},\ \bibinfo {pages} {050601} (\bibinfo {year} {2020})}\BibitemShut {NoStop}%
\bibitem [{\citenamefont {Nishiyama}\ and\ \citenamefont {Hasegawa}(2024{\natexlab{b}})}]{Nishiyama:2024:ExactQDAPRE}%
  \BibitemOpen
  \bibfield  {author} {\bibinfo {author} {\bibfnamefont {T.}~\bibnamefont {Nishiyama}}\ and\ \bibinfo {author} {\bibfnamefont {Y.}~\bibnamefont {Hasegawa}},\ }\bibfield  {title} {\bibinfo {title} {Exact solution to quantum dynamical activity},\ }\href {https://doi.org/10.1103/PhysRevE.109.044114} {\bibfield  {journal} {\bibinfo  {journal} {Phys. Rev. E}\ }\textbf {\bibinfo {volume} {109}},\ \bibinfo {pages} {044114} (\bibinfo {year} {2024}{\natexlab{b}})}\BibitemShut {NoStop}%
\bibitem [{\citenamefont {Nakajima}\ and\ \citenamefont {Utsumi}(2023)}]{Nakajima:2023:SLD}%
  \BibitemOpen
  \bibfield  {author} {\bibinfo {author} {\bibfnamefont {S.}~\bibnamefont {Nakajima}}\ and\ \bibinfo {author} {\bibfnamefont {Y.}~\bibnamefont {Utsumi}},\ }\bibfield  {title} {\bibinfo {title} {Symmetric-logarithmic-derivative {Fisher} information for kinetic uncertainty relations},\ }\href {https://doi.org/10.1103/PhysRevE.108.054136} {\bibfield  {journal} {\bibinfo  {journal} {Phys. Rev. E}\ }\textbf {\bibinfo {volume} {108}},\ \bibinfo {pages} {054136} (\bibinfo {year} {2023})}\BibitemShut {NoStop}%
\bibitem [{\citenamefont {Nishiyama}\ and\ \citenamefont {Hasegawa}(2024{\natexlab{c}})}]{Nishiyama:2024:OpenQuantumRURJPA}%
  \BibitemOpen
  \bibfield  {author} {\bibinfo {author} {\bibfnamefont {T.}~\bibnamefont {Nishiyama}}\ and\ \bibinfo {author} {\bibfnamefont {Y.}~\bibnamefont {Hasegawa}},\ }\bibfield  {title} {\bibinfo {title} {Tradeoff relations in open quantum dynamics via {Robertson}, {Maccone}-{Pati}, and {Robertson}-{Schr{\"o}dinger} uncertainty relations},\ }\href {https://doi.org/10.1088/1751-8121/ad79cd} {\bibfield  {journal} {\bibinfo  {journal} {J. Phys. A: Math. Theor.}\ }\textbf {\bibinfo {volume} {57}},\ \bibinfo {pages} {415301} (\bibinfo {year} {2024}{\natexlab{c}})}\BibitemShut {NoStop}%
\bibitem [{\citenamefont {Hasegawa}\ and\ \citenamefont {Nishiyama}(2024)}]{Hasegawa:2024:ConcentrationIneqPRL}%
  \BibitemOpen
  \bibfield  {author} {\bibinfo {author} {\bibfnamefont {Y.}~\bibnamefont {Hasegawa}}\ and\ \bibinfo {author} {\bibfnamefont {T.}~\bibnamefont {Nishiyama}},\ }\bibfield  {title} {\bibinfo {title} {Thermodynamic concentration inequalities and trade-off relations},\ }\href {https://doi.org/10.1103/PhysRevLett.133.247101} {\bibfield  {journal} {\bibinfo  {journal} {Phys. Rev. Lett.}\ }\textbf {\bibinfo {volume} {133}},\ \bibinfo {pages} {247101} (\bibinfo {year} {2024})}\BibitemShut {NoStop}%
\bibitem [{\citenamefont {Shiraishi}\ and\ \citenamefont {Saito}(2021)}]{PhysRevResearch.3.023074}%
  \BibitemOpen
  \bibfield  {author} {\bibinfo {author} {\bibfnamefont {N.}~\bibnamefont {Shiraishi}}\ and\ \bibinfo {author} {\bibfnamefont {K.}~\bibnamefont {Saito}},\ }\bibfield  {title} {\bibinfo {title} {Speed limit for open systems coupled to general environments},\ }\href {https://doi.org/10.1103/PhysRevResearch.3.023074} {\bibfield  {journal} {\bibinfo  {journal} {Phys. Rev. Res.}\ }\textbf {\bibinfo {volume} {3}},\ \bibinfo {pages} {023074} (\bibinfo {year} {2021})}\BibitemShut {NoStop}%
\bibitem [{\citenamefont {Yuto~Ashida}\ and\ \citenamefont {Ueda}(2020)}]{Ashida:2020:NonHermiteReview}%
  \BibitemOpen
  \bibfield  {author} {\bibinfo {author} {\bibfnamefont {Z.~G.}\ \bibnamefont {Yuto~Ashida}}\ and\ \bibinfo {author} {\bibfnamefont {M.}~\bibnamefont {Ueda}},\ }\bibfield  {title} {\bibinfo {title} {Non-hermitian physics},\ }\href {https://doi.org/10.1080/00018732.2021.1876991} {\bibfield  {journal} {\bibinfo  {journal} {Advances in Physics}\ }\textbf {\bibinfo {volume} {69}},\ \bibinfo {pages} {249} (\bibinfo {year} {2020})}\BibitemShut {NoStop}%
\bibitem [{\citenamefont {M{\o}lmer}\ \emph {et~al.}(1993)\citenamefont {M{\o}lmer}, \citenamefont {Castin},\ and\ \citenamefont {Dalibard}}]{Molmer:1993:MonteCarlo}%
  \BibitemOpen
  \bibfield  {author} {\bibinfo {author} {\bibfnamefont {K.}~\bibnamefont {M{\o}lmer}}, \bibinfo {author} {\bibfnamefont {Y.}~\bibnamefont {Castin}},\ and\ \bibinfo {author} {\bibfnamefont {J.}~\bibnamefont {Dalibard}},\ }\bibfield  {title} {\bibinfo {title} {{Monte} {Carlo} wave-function method in quantum optics},\ }\href {https://doi.org/10.1364/JOSAB.10.000524} {\bibfield  {journal} {\bibinfo  {journal} {J. Opt. Soc. Am. B}\ }\textbf {\bibinfo {volume} {10}},\ \bibinfo {pages} {524} (\bibinfo {year} {1993})}\BibitemShut {NoStop}%
\bibitem [{\citenamefont {Carmichael}(2009)}]{Carmichael:2009:QuantumTrajLecture}%
  \BibitemOpen
  \bibfield  {author} {\bibinfo {author} {\bibfnamefont {H.}~\bibnamefont {Carmichael}},\ }\href {https://link.springer.com/book/10.1007/978-3-540-47620-7} {\emph {\bibinfo {title} {An open systems approach to quantum optics: lectures presented at the Universit{\'e} Libre de Bruxelles, October 28 to November 4, 1991}}},\ Vol.~\bibinfo {volume} {18}\ (\bibinfo  {publisher} {Springer Science \& Business Media},\ \bibinfo {year} {2009})\BibitemShut {NoStop}%
\bibitem [{\citenamefont {Uzdin}\ \emph {et~al.}(2012)\citenamefont {Uzdin}, \citenamefont {G{\"u}nther}, \citenamefont {Rahav},\ and\ \citenamefont {Moiseyev}}]{Uzdin:2012:NonHermitianSL}%
  \BibitemOpen
  \bibfield  {author} {\bibinfo {author} {\bibfnamefont {R.}~\bibnamefont {Uzdin}}, \bibinfo {author} {\bibfnamefont {U.}~\bibnamefont {G{\"u}nther}}, \bibinfo {author} {\bibfnamefont {S.}~\bibnamefont {Rahav}},\ and\ \bibinfo {author} {\bibfnamefont {N.}~\bibnamefont {Moiseyev}},\ }\bibfield  {title} {\bibinfo {title} {Time-dependent hamiltonians with $100\%$ evolution speed efficiency},\ }\href {https://dx.doi.org/10.1088/1751-8113/45/41/415304} {\bibfield  {journal} {\bibinfo  {journal} {J. Phys. A: Math. Theor.}\ }\textbf {\bibinfo {volume} {45}},\ \bibinfo {pages} {415304} (\bibinfo {year} {2012})}\BibitemShut {NoStop}%
\bibitem [{\citenamefont {Impens}\ \emph {et~al.}(2021)\citenamefont {Impens}, \citenamefont {D'Angelis}, \citenamefont {Pinheiro},\ and\ \citenamefont {Gu{\'e}ry-Odelin}}]{Impens:2021:nonHermitianSpeedLimit}%
  \BibitemOpen
  \bibfield  {author} {\bibinfo {author} {\bibfnamefont {F.}~\bibnamefont {Impens}}, \bibinfo {author} {\bibfnamefont {F.~M.}\ \bibnamefont {D'Angelis}}, \bibinfo {author} {\bibfnamefont {F.~A.}\ \bibnamefont {Pinheiro}},\ and\ \bibinfo {author} {\bibfnamefont {D.}~\bibnamefont {Gu{\'e}ry-Odelin}},\ }\bibfield  {title} {\bibinfo {title} {Time scaling and quantum speed limit in non-hermitian hamiltonians},\ }\href {https://doi.org/10.1103/PhysRevA.104.052620} {\bibfield  {journal} {\bibinfo  {journal} {Phys. Rev. A}\ }\textbf {\bibinfo {volume} {104}},\ \bibinfo {pages} {052620} (\bibinfo {year} {2021})}\BibitemShut {NoStop}%
\bibitem [{\citenamefont {Thakuria}\ \emph {et~al.}(2023)\citenamefont {Thakuria}, \citenamefont {Srivastav}, \citenamefont {Mohan}, \citenamefont {Kumari},\ and\ \citenamefont {Pati}}]{Thakuria:2024:GQSL}%
  \BibitemOpen
  \bibfield  {author} {\bibinfo {author} {\bibfnamefont {D.}~\bibnamefont {Thakuria}}, \bibinfo {author} {\bibfnamefont {A.}~\bibnamefont {Srivastav}}, \bibinfo {author} {\bibfnamefont {B.}~\bibnamefont {Mohan}}, \bibinfo {author} {\bibfnamefont {A.}~\bibnamefont {Kumari}},\ and\ \bibinfo {author} {\bibfnamefont {A.~K.}\ \bibnamefont {Pati}},\ }\bibfield  {title} {\bibinfo {title} {Generalised quantum speed limit for arbitrary time-continuous evolution},\ }\href {https://dx.doi.org/10.1088/1751-8121/ad15ad} {\bibfield  {journal} {\bibinfo  {journal} {J. Phys. A: Math. Theor.}\ }\textbf {\bibinfo {volume} {57}},\ \bibinfo {pages} {025302} (\bibinfo {year} {2023})}\BibitemShut {NoStop}%
\bibitem [{\citenamefont {Nishiyama}\ and\ \citenamefont {Hasegawa}(2025)}]{Nishiyama:2024:NonHermiteQSLPRA}%
  \BibitemOpen
  \bibfield  {author} {\bibinfo {author} {\bibfnamefont {T.}~\bibnamefont {Nishiyama}}\ and\ \bibinfo {author} {\bibfnamefont {Y.}~\bibnamefont {Hasegawa}},\ }\bibfield  {title} {\bibinfo {title} {Speed limits and thermodynamic uncertainty relations for quantum systems with the non-{Hermitian} {Hamiltonian}},\ }\href {https://doi.org/10.1103/PhysRevA.111.012214} {\bibfield  {journal} {\bibinfo  {journal} {Phys. Rev. A}\ }\textbf {\bibinfo {volume} {111}},\ \bibinfo {pages} {012214} (\bibinfo {year} {2025})}\BibitemShut {NoStop}%
\bibitem [{\citenamefont {Bagrets}\ and\ \citenamefont {Nazarov}(2003)}]{Bagrets:2003:counting}%
  \BibitemOpen
  \bibfield  {author} {\bibinfo {author} {\bibfnamefont {D.~A.}\ \bibnamefont {Bagrets}}\ and\ \bibinfo {author} {\bibfnamefont {Y.~V.}\ \bibnamefont {Nazarov}},\ }\bibfield  {title} {\bibinfo {title} {Full counting statistics of charge transfer in coulomb blockade systems},\ }\href {https://doi.org/10.1103/PhysRevB.67.085316} {\bibfield  {journal} {\bibinfo  {journal} {Phys. Rev. B}\ }\textbf {\bibinfo {volume} {67}},\ \bibinfo {pages} {085316} (\bibinfo {year} {2003})}\BibitemShut {NoStop}%
\bibitem [{\citenamefont {Kir{\v{s}}anskas}\ \emph {et~al.}(2017)\citenamefont {Kir{\v{s}}anskas}, \citenamefont {Pedersen}, \citenamefont {Karlstr{\"o}m}, \citenamefont {Leijnse},\ and\ \citenamefont {Wacker}}]{Kirsanskas:2017:QmeQ}%
  \BibitemOpen
  \bibfield  {author} {\bibinfo {author} {\bibfnamefont {G.}~\bibnamefont {Kir{\v{s}}anskas}}, \bibinfo {author} {\bibfnamefont {J.~N.}\ \bibnamefont {Pedersen}}, \bibinfo {author} {\bibfnamefont {O.}~\bibnamefont {Karlstr{\"o}m}}, \bibinfo {author} {\bibfnamefont {M.}~\bibnamefont {Leijnse}},\ and\ \bibinfo {author} {\bibfnamefont {A.}~\bibnamefont {Wacker}},\ }\bibfield  {title} {\bibinfo {title} {{QmeQ 1.0}: An open-source {Python} package for calculations of transport through quantum dot devices},\ }\href {https://doi.org/10.1016/j.cpc.2017.07.024} {\bibfield  {journal} {\bibinfo  {journal} {Comput. Phys. Commun.}\ }\textbf {\bibinfo {volume} {221}},\ \bibinfo {pages} {317} (\bibinfo {year} {2017})}\BibitemShut {NoStop}%
\bibitem [{\citenamefont {van~der Wiel}\ \emph {et~al.}(2002)\citenamefont {van~der Wiel}, \citenamefont {De~Franceschi}, \citenamefont {Elzerman}, \citenamefont {Fujisawa}, \citenamefont {Tarucha},\ and\ \citenamefont {Kouwenhoven}}]{vanderWiel:2002:doubleQuantumDots}%
  \BibitemOpen
  \bibfield  {author} {\bibinfo {author} {\bibfnamefont {W.~G.}\ \bibnamefont {van~der Wiel}}, \bibinfo {author} {\bibfnamefont {S.}~\bibnamefont {De~Franceschi}}, \bibinfo {author} {\bibfnamefont {J.~M.}\ \bibnamefont {Elzerman}}, \bibinfo {author} {\bibfnamefont {T.}~\bibnamefont {Fujisawa}}, \bibinfo {author} {\bibfnamefont {S.}~\bibnamefont {Tarucha}},\ and\ \bibinfo {author} {\bibfnamefont {L.~P.}\ \bibnamefont {Kouwenhoven}},\ }\bibfield  {title} {\bibinfo {title} {Electron transport through double quantum dots},\ }\href {https://doi.org/10.1103/RevModPhys.75.1} {\bibfield  {journal} {\bibinfo  {journal} {Rev. Mod. Phys.}\ }\textbf {\bibinfo {volume} {75}},\ \bibinfo {pages} {1} (\bibinfo {year} {2002})}\BibitemShut {NoStop}%
\bibitem [{\citenamefont {Carrascal}\ \emph {et~al.}(2015)\citenamefont {Carrascal}, \citenamefont {Ferrer}, \citenamefont {Smith},\ and\ \citenamefont {Burke}}]{Carrascal:2015:HubbardDimer}%
  \BibitemOpen
  \bibfield  {author} {\bibinfo {author} {\bibfnamefont {D.~J.}\ \bibnamefont {Carrascal}}, \bibinfo {author} {\bibfnamefont {J.}~\bibnamefont {Ferrer}}, \bibinfo {author} {\bibfnamefont {J.~C.}\ \bibnamefont {Smith}},\ and\ \bibinfo {author} {\bibfnamefont {K.}~\bibnamefont {Burke}},\ }\bibfield  {title} {\bibinfo {title} {The {Hubbard} dimer: a density functional case study of a many-body problem},\ }\href {https://doi.org/10.1088/0953-8984/27/39/393001} {\bibfield  {journal} {\bibinfo  {journal} {J. Phys.: Condens. Matter}\ }\textbf {\bibinfo {volume} {27}},\ \bibinfo {pages} {393001} (\bibinfo {year} {2015})}\BibitemShut {NoStop}%
\bibitem [{\citenamefont {Wang}\ \emph {et~al.}(2024)\citenamefont {Wang}, \citenamefont {Rooney},\ and\ \citenamefont {Jiang}}]{Wang:2024:HubbardDQDCharacterization}%
  \BibitemOpen
  \bibfield  {author} {\bibinfo {author} {\bibfnamefont {W.}~\bibnamefont {Wang}}, \bibinfo {author} {\bibfnamefont {J.~D.}\ \bibnamefont {Rooney}},\ and\ \bibinfo {author} {\bibfnamefont {H.}~\bibnamefont {Jiang}},\ }\bibfield  {title} {\bibinfo {title} {Efficient characterization of a double quantum dot using the {Hubbard} model},\ }\href {https://doi.org/10.1063/5.0215622} {\bibfield  {journal} {\bibinfo  {journal} {J. Appl. Phys.}\ }\textbf {\bibinfo {volume} {136}},\ \bibinfo {pages} {044401} (\bibinfo {year} {2024})}\BibitemShut {NoStop}%
\bibitem [{\citenamefont {Risken}(1989)}]{Risken:1989:FPEBook}%
  \BibitemOpen
  \bibfield  {author} {\bibinfo {author} {\bibfnamefont {H.}~\bibnamefont {Risken}},\ }\href@noop {} {\emph {\bibinfo {title} {The {Fokker}--{Planck} Equation: Methods of Solution and Applications}}},\ \bibinfo {edition} {2nd}\ ed.\ (\bibinfo  {publisher} {Springer},\ \bibinfo {year} {1989})\BibitemShut {NoStop}%
\bibitem [{\citenamefont {Shiraishi}\ \emph {et~al.}(2016)\citenamefont {Shiraishi}, \citenamefont {Saito},\ and\ \citenamefont {Tasaki}}]{Shiraishi:2016:UniversalTradeoff}%
  \BibitemOpen
  \bibfield  {author} {\bibinfo {author} {\bibfnamefont {N.}~\bibnamefont {Shiraishi}}, \bibinfo {author} {\bibfnamefont {K.}~\bibnamefont {Saito}},\ and\ \bibinfo {author} {\bibfnamefont {H.}~\bibnamefont {Tasaki}},\ }\bibfield  {title} {\bibinfo {title} {Universal trade-off relation between power and efficiency for heat engines},\ }\href {https://doi.org/10.1103/PhysRevLett.117.190601} {\bibfield  {journal} {\bibinfo  {journal} {Phys. Rev. Lett.}\ }\textbf {\bibinfo {volume} {117}},\ \bibinfo {pages} {190601} (\bibinfo {year} {2016})}\BibitemShut {NoStop}%
\end{thebibliography}
\end{document}